# A Review on P2P Video Streaming

**Sabu M. Thampi**
*Indian Institute of Information Technology and Management – Kerala (IIITM-K), India*
smthampi@ieee.org

The main objective of this article is to provide an overview of P2P based Video-on-Demand and live streaming services. The article starts with an introduction to media streaming and its simplified architecture. Various solutions offering video streaming in the context of widespread usage of Internet are discussed. This is followed by a short introduction to P2P networks and its applications. A broad discussion on various P2P streaming schemes and P2P streaming applications are the main focus of this chapter. Finally, the security issues and solutions for P2P video streaming are discussed briefly.

## Introduction

Video has been an important media for communications and entertainment for many decades. Movie is a form of entertainment that enacts a story by screening a series of images giving the delusion of continuous movement. The trick was already known in second-century China, but remained inquisitiveness up to the end of the 19th century. The invention of motion picture camera around 1888 allowed the individual component images to be captured and stored on a single reel. For the first time, this has made possible the process of recording scenes in an automatic manner. Further to that, a hasty transformation occurred with the development of a motion picture projector to enlarge these moving picture shows onto a screen for an entire audience. Television broadcasting after its invention in 1928 has attracted billions of people from different part of the world to watch both live events and recorded videos simultaneously through their television sets. People moved from newspaper and radio to the more immersive experience of television as their primary source of entertainment and as a way to receive important information and news about the world [1]. For most of the twentieth century, the only ways to watch television were through over-the-air broadcasts and cable signals.

A third boost in the popularity of moving pictures came at the end of the 20th century with the invention of the Internet and of the World Wide Web. Web browsing and file transfer are the dominant services provided through the Internet. However, these kinds of service providing information about text, pictures and document exchange are no longer satisfied the demand of clients. Following the success of conventional radio and television broadcasting, research has been carried out into ways of delivering live media over the Internet to a personal computer. As a result, people have experimented with transmitting various multimedia data such as sound and video over the Internet. All multimedia content were distributed no differently than any other ordinary files such as text files and executable files. They were all transmitted as "files' using file downloading protocols such as ftp and http. The full file transfer, in the download mode, can often suffer unacceptably long transfer times, which depend on the size of the media file and the bandwidth of the transport channel. For example, if downloaded from http://www.mp3.com, an MP3 audio file encoded at 128 kbit/s and of 5 minutes duration will occupy 4.8 MB of the user's hard disk. Using a 28.8k dial-up modem, it would take roughly 40 minutes to download the whole file [2]. As a result, an audio file might take more real-time to download than the length of the audio being played. Video, which carries much more information than audio, entailed even longer download times [3]. Furthermore, there was no way for the users to "peek" into the content to see if it is the video they would like to watch. This was often inconvenient for the users due to a long waiting time and a large amount of wasted resources when the content of the video turned out to be something they were not interested in [4].

Internet evolves and operates basically without a central coordination, the lack of which was and is vitally important to the rapid escalation and evolution of Internet. However, the lack of management in turn makes it very difficult to guarantee proper performance and to deal systematically with performance issues. Meanwhile, the available network bandwidth and server capacity continue to be besieged by the mounting Internet utilization and the accelerating escalation of bandwidth demanding content. As a result, Internet service quality perceived by customers is largely unpredictable and inadequate [5].The current Internet is inherently a packet-switched network that was not designed to handle continuous time-based traffic such as audio and video. The Internet only provides best-effort services and has no guarantee on the quality of service (QoS) for multimedia data transmission [6].

Recent advances in digital technologies such as high-speed networking, media compression technologies and fast computer processing power, have made it feasible to provide real-time multimedia services over the Internet. Real-time multimedia, as the name implies, has timing constraints. For example, audio and video data must be played out continuously. If the data does not arrive in time, the play out process will pause, which is annoying to human ears and eyes. Real-time transport of live video or stored video is the predominant part of real-time multimedia. *Streaming* is an enabling technology for providing multimedia data delivery among clients in various multimedia applications on the Internet. With this technology, the client can playback the media content without waiting for the entire media file to arrive. Thus, streaming allows real-time transmission of multimedia over the net. Internet streaming media changed the Web as we knew it-- changed it from a static text- and graphics-based medium into a multimedia experience populated by sound and moving pictures [7]. Websites such as *You Tube*, provide media content to millions of viewers. American National Standard for Telecommunications defines *streaming* as " *a technique for transferring data (usually over the Internet) in a continuous flow to allow large multimedia files to be viewed before the entire file has been downloaded to a client's computer*" [8]. The basic idea of video streaming is to split the video into parts, transmit these parts in succession, and enable the receiver to decode and playback the video as these parts are received, without having to wait for the entire video to be delivered. Thus, streaming enables near instantaneous playback of multimedia content in spite of their sizes. Streaming media utilizes a very old concept called *buffering* to make feasible the playback of multimedia content as it is being downloaded. A buffer clasps a pool of content sufficiently large to stabilize the bumps in playback that may be caused by transitory server slowdown or network overcrowding.

Streaming diminishes the storage space and permits users to stop receiving the stream, if not interesting or satisfactory, before the entire file is downloaded. Streaming allows live and pre-recoded content to be distributed. Live streaming captures audio/video signals from input devices (e.g. microphone, video camera), encodes the signals using compression algorithms (e.g. MP3, MPEG-4), and distributes them in real-time. Typical application of live streaming includes surveillance, broadcasting of special events, and distribution of information that have the prime importance in real-time delivery. In live streaming, the server side has the control over the selection of the distribution content and the timing of their streaming. The user involvement is typically limited to joining and leaving the running streaming sessions. Pre-recorded or stored streaming distributes pre-encoded video files stored at a media server. Sample applications include multimedia archival retrievals, news clip viewing, and distance learning through which students attend classes on-line by viewing pre-recorded lectures [4].

With the rise of broadband Internet connections, end users became able to receive video of acceptable quality on their home computers. Broadband has achieved mass-market penetration in several countries. According to world's leading information technology research and advisory company - Gartner, worldwide consumer broadband connections will grow from 323 million connections in 2007 to 580m in 2013. This ensures that a large number of consumers will have sufficient bandwidth to receive streaming video and audio in the near future. Now streaming media is poised to become the de facto global media broadcasting and distribution standard, incorporating all other media, including television, radio, and film. According to an industry study [11], there were more than, 60 million people listening to or watching streaming media each month, 58 US TV stations performing live webcasting, 34 offering on-demand streaming media programs, and 69 international TV webcasters. The study also finds that 6000 hours of new streaming programming are created each week. The market for streaming content has grown substantially in Europe. For instance, the BBC, which reaches an audience of over 1 million a month, estimates that its streaming audience size is growing by 100 percent every four months. One of the leading French streaming sites, CanalWeb, boasts over 450,000 unique viewers per month, with video content watched for an average of 12 minutes. In the UK, RealNetworks estimates that 500,000 users downloaded its player from the Big Brother Web site

(www.bigbrother2000.com). Big Brother UK reports it was serving at least 6,000 simultaneous streams, and 1.5 million per day. Market research firm NetValue reports that the average viewing time for these streams was 25 minutes. RealPlayer users are an increasingly international group, totaling over 48 million regular users, with approximately one-third of downloads/registrations now originating outside North America [11].

## Architecture for Video Streaming

Figure 1 shows architecture for video streaming and it is divided into six areas as follows: *media compression, application-layer QoS control, media distribution services, streaming servers, media synchronization at the receiver side,* and *streaming media protocols*.

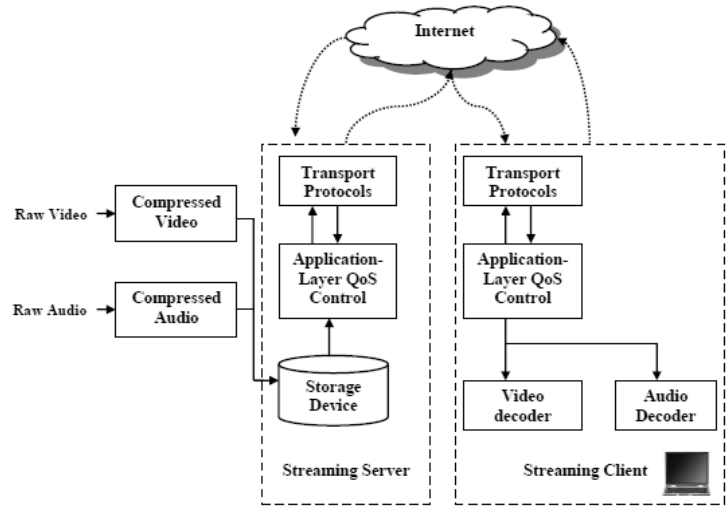

**Figure 1: Video Streaming Architecture** [12]

*Media compression:* The large volume of raw multimedia data imposes a stringent bandwidth requirement on the network. Hence, for achieving better transmission efficiency, compression is widely employed. While video needs superior bandwidth requirements (56 Kbps-15 Mbps) than audio (8 Kbps-128 Kbps) and loss of audio is more infuriating to human than video, audio is given higher priority for transmission in a multimedia streaming system. For this reason, only video will be used for alteration so as to meet the QoS requirements [6]. In Figure 1, raw video and audio data are pre-compressed by video compression and audio compression algorithms and then saved in storage devices. Video compression is accomplished by utilizing the resemblances or redundancies that subsists in a normal video signal. Video compression reduces the irrelevancy in the video signal by only coding video features that are perceptually important [13]. Video compression follows a standard for multimedia contents that encodes the content with a specific play rate. There are two major groups which define the video encoders: ITU (International Telecommunications Union) and ISO (International Standards Organization). ITU-T group (Telecommunication Standardization Sector of the International Telecommunications Union) defines the H.26x video formats whereas the ISO group defines the formats which have materialized from committees of the Moving Pictures Experts Group: MPEG-x. The MPEG-4 standard is commonly designed for streaming media and CD distribution, video conversion and broadcast television. MPEG-4 includes numerous features of MPEG-1, MPEG-2 and other associated standards. H.264 is also known as MPEG-4 part 10 or AVC (Advanced Video Coding). Big Internet players like Google/ You Tube or Apple Tunes are founded on this standard.

*Application-layer QoS control:* Upon the client's request, a *streaming server* retrieves compressed video/audio data from storage devices and then the *application-layer QoS control* module adapts the video/audio bit-streams according to the network status and QoS requirements. The application-layer QoS control involves *congestion control* and *error control* which are implemented at the application layer. The former is used to determine the transmission rate of media streams based on the estimated network bandwidth while the latter aims at matching the rate of a precompressed media bit streams to the target rate constraint by using filtering [12].

Typically, for streaming video, congestion control takes the form of *rate control*. Rate control attempts to minimize the possibility of network congestion by matching the rate of the video stream to the available network bandwidth. Based on the place where rate control is taken in the system, rate control can be categorized into three types: *source-based, receiver-based and hybrid-based*. With the source-based rate control, only the sender (server) is responsible for adapting the transmission rate. In contrast, the receiving rate of the streams is regulated by the client in the receiver-based method. Hybrid-based rate control employs the aforementioned schemes at the same time, i.e. both the server and client are needed to participant in the rate control. Typically, the source-based scheme is used in either unicast or multicast environment while the receiver-based method is deployed in multicast only [6].

The function of error control is to improve video presentation quality in the presence of packet loss. Error control mechanisms include *Forward Error Correction (FEC), retransmission, error-resilient encoding and error concealment*. With FEC scheme, the received packets at the receiver end are FEC decoded and unpacked, and the resulting bit stream is then input to the video decoder to reconstruct the original video. *Error-resilient encoding* is executed by the source to enhance robustness of compressed video before packet loss actually happens. Even when an image sample or a block of samples are missing due to transmission errors, the decoder can try to estimate them based on surrounding received samples, by making use of inherent correlation among spatially and temporally adjacent samples, such techniques arc known as *error concealment* techniques [14].

*Media distribution services:* After the adaptation by *application-layer QoS control* module, the *transport protocols* packetize the compressed bit-streams and send the video/audio packets to the Internet. Packets may be dropped or experience excessive delay inside the Internet due to congestion. In addition to the application-layer support, adequate *network* support is necessary to reduce transport delays and packet losses. The network support involves *network filtering, application-level multicast and content replication (caching)*. Network filtering maximizes video quality during network congestion. The filter at the video server can adapt the rate of video streams according to the network congestion status. The *application-level multicast* provides a multicast service on top of the Internet. These protocols do not modify the network infrastructure; instead they employ multicast forwarding functionality solely at end-hosts. *Content replication improves scalability of the media delivery system.*

*Streaming servers:* Streaming servers play an important role in providing streaming services. To offer superiority streaming services, streaming servers are required to process multimedia data in real time, support VCR like functions and retrieve media components in a synchronous fashion. A streaming server generally waits for a Real Time Streaming Protocol (RTSP) request from the viewers. When it gets a request, the server looks in the appropriate folder for a hinted media of the requested name. If the requested media is in the folder, the server streams it to the viewer using RTP (Real-time Transport Protocol) streams.

*Media synchronization at the receiver side:* With media synchronization mechanisms, the application at the receiver side can present various media streams in the same way as they were originally captured. An example of media synchronization is synchronizing the movements of a speaker's lips with the sound of his speech.

*Protocols for streaming media:* Streaming protocols provide means to the client and the server for services negotiation, data transmission and network addressing. According to the functionalities, the protocols directly related to Internet streaming video can be classified as *network-layer protocol, transport protocol* and *session control protocol*.

Network-layer protocol provides basic network service support such as network addressing. The IP serves as the *network-layer protocol* for Internet video streaming. *Transport protocol* provides end-to-end network transport functions for streaming applications. Transport protocols include UDP, TCP, RTP, and real-time control protocol (RTCP). RTP and RTCP are upper-layer transport protocols implemented on top of UDP/TCP. UDP and TCP protocols support such functions as multiplexing, error control, congestion control, or flow control. RTP is a data transfer protocol. RTCP provides QoS feedback to the participants of an RTP session. *Session control protocol* defines the messages and procedures to control the delivery of the multimedia data during an established session.

RTSP and the session initiation protocol (SIP) are such session control protocols. RTSP is a protocol for use in streaming media systems which allows a client to remotely control a streaming media server, issuing VCR-like commands. It also allows time-based access to files on a server. SIP is a session protocol which can create and terminate sessions with one or more participants. It is mainly designed for interactive multimedia application, such as Internet phone and video conferencing [6].

# Existing Streaming Networks

There are three important means in which a streaming service may be offered over the Internet. The first approach employs caching and replication for the web based distribution for small amount of streaming media. For a large scale service, streaming content is distributed through a Content Delivery Network (CDN) which perks up the scalability of Web based content sharing. The second method is to use a network specifically designed for the distribution of streaming content. A number of networks have been proposed that are specialized in on-demand delivery of video streams. These networks are called as *On-demand Multimedia Streaming Networks*. The third option - live streaming systems allow clients to simultaneously watch a number of Television stations through the broadband Internet connectivity available at their homes.

## *Web Based Distribution*

Web based distribution is the most frequently used technique to serve small streaming content. As the Internet has become a vital part of daily life, hundreds of millions of users currently connect to the Internet. Due to client-server based computing model, Web based content distribution architecture suffers from server overloading when a large number of user requests arrive. Hence, appropriate schemes are required to manage the server loads effectively. *Content caching* and *replication techniques* direct the workload away from possibly overloaded origin Web servers to deal with Web performance and scalability from the client side and the server side, respectively [15]. Content Delivery Networks (CDN) is another approach being widely employed to perk up Internet service quality.

*Caching* stores a copy of data close to the data consumer to allow faster data access than if the content had to be retrieved from the origin server. For pre-recorded content, a streaming media-caching server can fetch and store entire contents for a user. When other users request the similar content, the cache can deliver the stream directly out of its local storage. Web caching lessens the access latency, saves CPU cycle of a Web server, and reduces the network bandwidth usage. However, it is usually considered not an excellent solution for streaming video content as caching of a video stream requires a very large buffer space [16]. *Replication* creates and maintains distributed copies of content under the control of content providers. This is obliging because client requests can then be sent to the adjacent and least loaded server. Several web sites replicate their content at multiple servers with the intention of reducing the load on the originating server. Replication also provides server redundancy in case of server and network failures. On the other hand, due to the unique nature of the WWW, its massive user community, document multiplicity, and access patterns replication seems to be not able to stand up fully to all of its conceptual promises with respect to latency and bandwidth reduction [17].

A *Content Delivery Network (CDN)* replicates content from the origin server to cache servers (also called replica servers), spread across the globe. Content requests are directed to the cache server closest to the user, and that server delivers the requested content. As a result, users get greater speed and higher quality**.** There are two general approaches for building CDNs: *overlay* and *network* approach.

In the *overlay approach*, application-specific servers and caches at several places in the network handle the distribution of specific content types such as streaming media. Most of the commercial CDN providers such as Akamai and Limelight Networks follow the overlay approach for CDN organization. The core network components such as routers and switches play no active role in content delivery. Akamai system has more than 12,000 servers in over 1,000 networks. In the *network approach*, the network components including routers and switches are equipped with code for identifying specific application types and for forwarding the requests based on predefined policies. Examples of this approach include devices that redirect content requests to local caches or switch traffic to specific servers, optimized to serve specific content types [18]. Besides increased server capacity and resiliency, a CDN gives controlled load balancing and enhanced content accessibility. Operating servers in various locations creates several technical challenges, including how to direct user requests to suitable servers, how to manage failures, how to monitor and control the servers, and how to update software across the system [19, 20]. The amount of load the network can manage is preset by the overall CDN capacity. Special events and programs frequently produce additional demands than what the network can handle in a short period of time and CDN will not be able to bear those excess demands. At those locations where demands becomes high, a suitable mechanism that permits dynamic addition and removal of replica servers is required. Accordingly, in order for a CDN to be really successful, a large number of replication servers must be set up throughout the Internet. Such an arrangement may not be possible by small organizations [4].

## *On-Demand Multimedia Streaming*

Video on demand (VoD) also known as *on-demand video streaming* is a great way of viewing films and television programs. VoD service enables immediate distribution of video streams to users, from the beginning of the content, regardless of the time at which the service request arrives in relation to other on-going streaming sessions. Typically, these video files are stored in a set of central video servers, and distributed through high speed communication networks to geographically-dispersed clients. Upon receiving a client's service request, a server delivers the video to the client as an isochronous video stream. VoD has become an extremely popular service in the Internet. For example, YouTube, a video-sharing service which streams its videos to users' on-demand, has more than 20 million views a day with a total viewing time of over 10,000 years to date. Other major Internet VoD publishers include MSN Video, Google Video, Yahoo Video, CNN, and a plethora of copycat YouTube sites [21]. VoD wipes out the necessity to go to your video store to buy films and offers access to a large collection of material. With VoD, users will have the flexibility of choosing the content as well as scheduling the program they desire to watch [22].

There are two major ways to implement the VoD architecture: *centralized architecture and distributed architecture*. In the centralized architecture clients are directly connected to the video server through the network as shown in figure 2. A video server has access to the video content storage and is responsible for the delivery of the video content in uninterrupted streams. Even though centralized VoD systems are simple to manage, the major problem of this architecture lies in the poor scalability as the service capacity is well defined by server limitations. A system expansion may lead to huge costs in resources increment. When the number of clients increase, the number of streams needed may be enormous resulting additional channel bandwidth. The performance of centralized VoD systems can be improved by adding local servers. The local sites do not maintain media archives; however they can store popular movies in their video buffers. The contents of the buffers can be delivered to clients more quickly without accessing the central server. Videos that are not buffered at local sites can be delivered to clients from the central archive when they are requested.

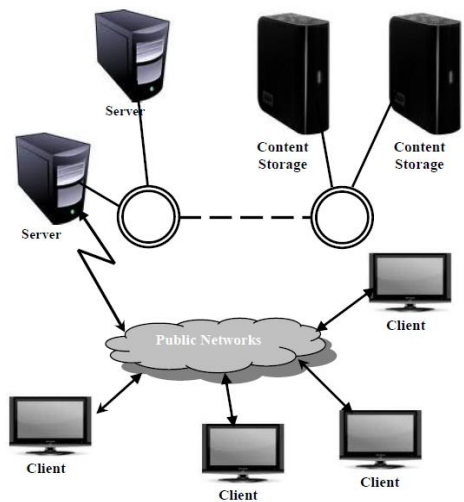

**Figure 2: VoD Centralized Architecture**

In *distributed architecture*, multiple video servers are distributed throughout the network infrastructure. Each video server controls and manages a subset of the content storage and is responsible for a subset of the video streams. Figure 3 shows a typical layout for distributed architecture. Ideally, all the popular content is replicated at the video servers connected to each exchange. This significantly decreases traffic between the servers and as a result settles down the bandwidth requirements between the main hubs. If a local server does not have a requested video title, it searches through a list of all the video servers which have that title and picks the one with the least network load. Distributed architecture is a viable option as it relaxes the bandwidth requirements on the network.

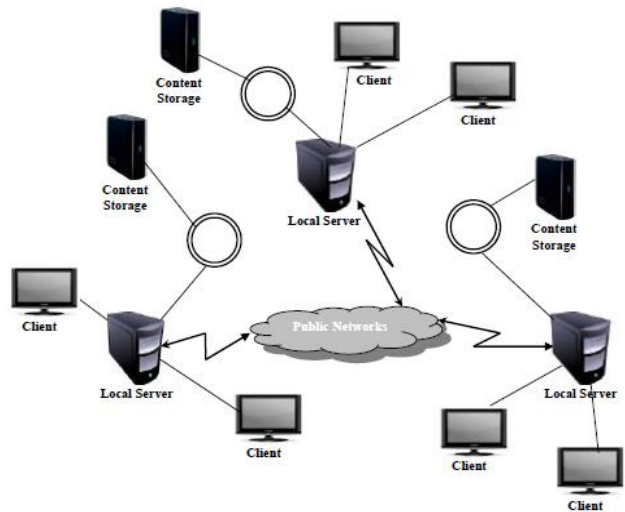

**Figure 3: VoD Distributed Architecture**

In VoD system, a popular content can attract a large number of viewers, who performs request asynchronously. As there are dedicated channels to service each user request, the bandwidth requirements are increased significantly as more and more streams are requested. The fundamental challenge of VoD service is how to meet the on-demand expectation of users without consuming a large amount of bandwidth at the content server. A number of schemes have been proposed that focus on the efficient bandwidth usage of the content server. A common thread in all schemes is the use of *multicasting*. Since a user can not watch the video at once, broadcasting protocols can only provide near VoD service. Multicast transmission sends exactly one copy of the stream, not over the whole network but only down the branches of the network where one or more viewers are tuned in. In this way, the available network bandwidth can be used more efficiently. For multicast in a VoD system, there is a possibility that more than one user requests the same video at the same time. The probability of more than one user requesting the same video will be high if the number of videos available is small when compared to the number of users. Even if there is a large video archive, there will be a set of popular videos which are requested by many users, thus increasing the chance of multicasting. If a requested video is multicast, all the users in the multicast group will be served by one channel thus saving the network bandwidth [22]. Some multicasting schemes propose how to provide efficient and practical multicasting while others assume the availability of multicasting to all participating users. Multicasting is mainly implemented in three ways: *IP multicast, overlay network based multicast and application layer multicast*.

*IP multicast* implements the service at the IP layer and offers efficient group communication. IP multicast requires fairly sophisticated router software that allows the server to replicate streams as required by the clients. The user of a multicast has no control over the media presented. Like in *broadcast*, the choice is simply to watch or not to watch. The user's host communicates with the nearest router to get a copy of the stream. Four classes of IP multicasting approaches have been proposed to overpass the gap between synchronous IP multicast and asynchronous VoD streaming: *batching, patching, periodic broadcasting and merging* [23].

The basic idea of 'batching' is to delay the requests for the different videos for a certain amount of time (batching interval) so that more requests for the same video arriving during the current batching interval may be serviced using the same stream. Thus, requests which are made by many different viewers for the same video can share a common video stream if these requests are spaced closely enough [24]. Batching can only be used with popular videos since unpopular videos are unlikely to receive multiple requests during the delay interval. While clients' requests are not instantly granted, the batching technique in fact offers a near-VoD service, but not a true VoD service.

In 'patching scheme', an existing multicast can expand dynamically to serve new clients. Most of the communication bandwidth of the server is organized into a set of logic channels and each is capable of transmitting a video at the playback rate. The remaining bandwidth of the server is used for control messages such as service requests and service notifications [25]. A channel is either a regular channel in which the server multicasts the entire video or a patching channel in which the server multicasts only the leading part of the video. When a client requests a video from the server, the server instructs the client to download from a regular channel and a patching channel. The client exits the patching channel after it downloads the leading part of the video but remains in the regular channel until the end of the video [23]. Patching is very simple and doesn't require any specialized hardware. Since all requests can be served immediately, the clients experience no service delay and true video on-demand can be achieved. Patching is very effective in reducing the bandwidth and storage requirements if the number of requests from the users is within a certain limit. Beyond that, patching looses its competence as it results in starting multiple patches of the same video and augments the bandwidth needs.

The idea behind 'periodic broadcasting scheme' is to divide the video into a series of segments and broadcast each segment periodically on dedicated server channels. Clients wait for the beginning of the first segment, and download the data of the next segment while watching the current segment. User waiting time is usually the length of the first segment [26]. In [26] divides the Periodic broadcasting protocols into three groups: *Pyramid Broadcasting, Harmonic Broadcasting and Hybrid broadcasting. Pyramid-like schemes* such as [27] and [28] have increasing size segments and equal bandwidth channels. The segment size of the videos in this protocol follows a geometrical series and different videos are mingled together in each logical channel. In *pyramid broadcasting,* the system requires that the video data be transferred at a rate much higher than it is consumed to provide on time delivery of the videos. In this scheme, video segments are of geometrically increasing sizes, and the server network bandwidth is evenly divided to periodically broadcast one segment in a separate channel. This solution requires expensive client machines with enough bandwidth to cope with the high data rate on each broadcast channel. *Harmonic-like schemes* such as [29] have equal size segments and decreasing bandwidth channels. They divide the video into equal size segments and transmit them in logical channels of decreasing bandwidth. This requires much less server bandwidth than pyramid broadcasting protocols. A new family of the 'hybrid broadcasting protocol' includes Pagoda broadcasting [30] and New Pagoda [31] broadcasting schemes. These protocols are hybrid of pyramid-based protocols and harmonic-based protocols. They partition each video into fixed size segments and map them into a small number of data streams of equal bandwidth and use time division multiplexing to ensure that successive segments of a given video are broadcast at the proper decreasing frequencies. The result is that they do not require significantly more bandwidth and at the same time do not use more logical streams.

The common issue among batching and patching is that they require twice or more bandwidth at the user system than the nominal playback rate since users must establish multiple streaming sessions concurrently. They also require a substantial amount of disk space in order to store segments from one of the streams while the other is being played out. In addition, they all assume the availability of multicasting capability at all participating nodes. In the '*stream merging scheme'* [32] the key idea is to encode the media at a bit rate just slightly less than the client receives bandwidth. The receive bandwidth that is left unused during viewing is used to perform near-optimal hierarchical stream merging. This technique has been shown to be highly effective in reducing server bandwidth. However, it may take a long time to complete the merging process or may never be realized when the time gap is large between two sessions. It also requires an encoder/decoder system that dynamically changes the rate of stream.

Centralized video-on-demand systems such as CNN Pipeline, YouTube and Uitzending gemist have drawbacks due the limited scalability of these systems. Batching and patching would increase the scalability of these systems a lot, however there is no support for broadcasting or multicasting in the Internet backbone. IP Multicast requires routers to maintain per group state. However, very few routers on the Internet can support IP multicast. Overhauling the Internet with IP multicast capable routers is a task considered not feasible in the near future. The routing and forwarding table at the routers now need to maintain an entry corresponding to each unique multicast group address. This increases the overheads and complexities at the routers. Another issue is that there is a lack of experience with additional mechanisms like reliability and congestion control on top of IP Multicast, which makes the ISPs wary of enabling multicasting at the network layer [33]. For these and other reasons, researchers have looked in other ways to achieve an efficient and effective group communication such as *overlay network based multicasting* and a*pplication layer multicast.*

In *overlay network based multicasting*, a network dedicated for the purpose of multicasting is created on top of existing IP network. Only those routers, i.e. overlay nodes that are equipped with multicasting functionality participate in multicast specific service; other routers simply forward packets in multicast sessions as regular unicast flows. An example of overlay multicast implementation is OMNI - the Overlay Multicast Network Infrastructure [34], which offers an overlay architecture for media streaming applications.

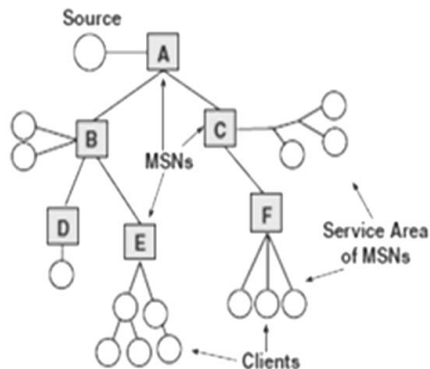

**Figure 4: OMNI Architecture [33]**

In OMNI, service providers deploy Multicast Service Nodes (MSNs) that run the routing and forwarding of information to a group of clients. OMNI follows a two-tier approach to overlay multicast (Figure 4). The lower tier contains a set of service nodes that are distributed throughout a core network infrastructure such as the Internet. The lower tier provides data distribution services to any host connected to an OMNI node. An end-host subscribes with a single OMNI node to receive multicast data service. The OMNI nodes organize themselves into an overlay which forms the multicast data delivery backbone. For the second layer, the data delivery path from the OMNI nodes to its clients is independent of the data delivery path used in the overlay backbone. This path can be built using network layer multicast, application-layer multicast, or a set of unicast paths [35]. The strengths of overlay network based multicasting include ability to deploy a large-scale multicast network without needing to upgrade all IP routers, support virtually unlimited number of multicast groups, and provide a practical solution for the deployment of group communication infrastructure on the Internet. However, it typically requires semi-permanently installed overlay nodes that will remain in service for an extended period of time or at least for the duration of the multicast session. For this reason, it is difficult to construct and maintain such a network within an environment where network nodes are highly dynamic, such as Ad-Hoc networks and Peer-to-Peer (P2P) networks.

Owing to the drawbacks presented by 'overlay multicast' and the slow deployment of 'IP multicast' technology on the global Internet, an application layer solution has been adopted; this approach is referred to as *Application Layer Multicast (ALM).* In ALM, the multicasting functionality is implemented at the application layer. ALM protocols do not change the network infrastructure; instead they employ multicast forwarding functionality exclusively at end-hosts. Unlike network layer multicast where data packets are replicated at routers inside the network, in application layer multicast data packets are replicated at end-hosts. In this multicast strategy, group membership, multicast tree construction and data forwarding are solely controlled by participating end hosts; thus, it does not require the support of intermediate nodes such as routers or dedicated servers. The P2P approach has ALM premises [23].

*Live Video Streaming*

Internet streaming technology also brings in more interesting applications such as transmission of traditional TV content in a much more flexible manner. Due to cost considerations, conventional TV networks normally offer channels only if there are enough user bases. For example, a TV network may be willing to offer Hindi programs in New York City where a large Indian population live, but not in many other parts of the country [36]. The introduction of live streaming services enables users to watch several TV channels through the Internet simultaneously. In live streaming, video streams are being generated at the same time as it is being downloaded and viewed by the clients. So, we are dealing with the distribution of a file of unknown and unpredictable length in which the data are only available for a small period of time. In this case, the most important challenge is the play-out delay, that is to say the time elapsing between the content production and its play-out. The lone action that the client should be able to carry out is to switch channels. The end-user experience is similar to a live TV broadcast as all of the users will intend to watch the most recently generated content. The user requires a download speed not less than equal to the playback speed if data loss is to be evaded. The popular live video streaming service is *Internet Protocol Television (IPTV)*. With the extensive acceptance of broadband residential access and the progress of video compression technologies, IPTV may be the next popular Internet application [37].

IPTV is a system where a digital television service is delivered using Internet Protocol over a network infrastructure, which may comprise delivery by a broadband Internet connection. So, IPTV offers digital television services over Internet Protocol (IP) for residential and business users at a lesser cost. The official definition approved by the International Telecommunication Union focus group on IPTV (ITU-T FG IPTV) is as follows: *"... multimedia services such as television/video/audio/text/graphics/data delivered over IP based networks managed to provide the required level of quality of service and experience, security, interactivity and reliability."* IPTV also makes it easier for users to access ostracized video on demand content, such as a well-known movies decades ago which is no longer offered in any important TV channels [36].

IPTV is a union of computing, communication, and content, as well as an amalgamation of broadcasting and telecommunication technologies. IPTV enables triple play of *voice, data and video*. The triple-play idea is that clients can subscribe to one service that offers voice, data, and video - all three brought into the home or office over one line, and by one service provider. The use of IP as a video delivery mechanism is omnipotent. An IPTV service system does not change the structure of content and channel production of the original Television Network. However, it just amends the controlled mode of transmission, i.e. it makes use of pure IP signaling to change channels and control other functions. In this fashion, selection space of content has been significantly expanded for user [38]. IPTV has a different infrastructure from TV services, which make use of a push metaphor in which the entire content is pushed to the clients. IPTV has two-way interactive communications between operators and users, for example, streaming control functions such as pause, forward, rewind, and so on, which traditional cable television services lack [39].

A typical IPTV system consists of four main components, as shown in Figure 5 [40]. The video headend (VH) captures all programming content, including linear programs and VoD content. The VH receives the content through satellite or terrestrial fiber networks. The VH also is responsible for encoding the video streams into MPEG-2 or MPEG-4 formats. VH encapsulates the video streams into a transport format and are sent to the core network (CN), using IP multicast or IP unicast. The CN groups the encoded video streams into their respective channels. The CN is unique to the service provider and often includes equipment from multiple vendors. At this stage, IPTV traffic can be protected from other Internet data traffic to guarantee a high level of QoS. The broadband remote access server (BRAS) is responsible for maintaining user policy management, such as subscriber authentication and accounting, IP address assignment, service advertisement etc. In the reverse direction, traffic from multiple end users is aggregated and routed to the core network by digital subscriber line access multiplexers (DSLAMs). The Home Network connects both the home computer(s) and the IP TV Set-Top Boxes (STB) to a broadband service to offer the data, voice, and video services in subscribing homes [40]. The STB converts a scrambled digital compressed signal into a signal that is sent to the TV.

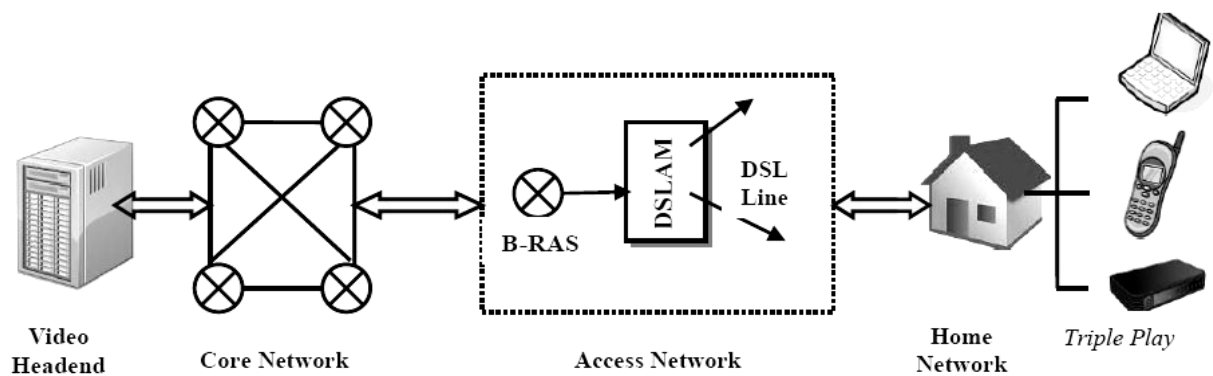

**Figure 5: IPTV System**

Globally, many of the world's major telecom providers are exploring IPTV as a new revenue opportunity from their existing markets. Two major U.S. telecommunication companies, AT&T and Verizon, have invested significantly to replace the copper lines in their networks with fiber optic cables for delivering many IPTV channels to residential customers [41]. The world's leading markets for IPTV for now are France, South Korea, Hong Kong, Japan, Italy, Spain, Belgium, China, Switzerland and Portugal. TV2 Sputnik is an IPTV service provider that uses the public Internet for content distribution. It is offered by one of the public service broadcasters in Denmark. Optimal Stream is an IPTV service provider, which delivers IPTV to Danish households over the public Internet. The United Kingdom launched IPTV early but has been slow to grow. IPTV is just beginning to grow in Central and Eastern Europe, now it is growing in South Asian countries such as Sri Lanka and especially India. Major vendors for IPTV in India include UTStarcom, Alcatel Lucent, SeaChange, Harmonic, Cisco, Irdeto, Harris, Viaccess, NDS, Conax, Verimatrix, Oracle, and Sun Microsystems. In [42] projects that consumer IP traffic will grow annually at 57%, driven and dominated by video traffic.

According to Multimedia Research Group (MRG) the number of global IPTV subscribers will grow from 44 million at the end of 2010 to 111.5 million in 2014, a compound annual growth rate of 26%. The forecast shows that Europe will be the regional leader, with 42% of the worldwide IPTV subscribers total in 2014, maintaining its lead mostly because of the sheer number of large Tier-1 Service Providers, and because of continued strong IPTV growth in some countries. According to a recent study by CISCO, the global IP traffic will continue to be dominated by video, exceeding 91 percent of global consumer IP traffic by 2014. The study expects that together all types of file-sharing traffic will nearly triple by 2014, still accounting for 27% of all Internet traffic. Internet video is predicted to account for 46% of all traffic in the same year.

## Failure of Traditional Streaming Techniques

The traditional client-server based streaming provides good performance and high availability rates if number of clients are limited. However, the deployment and maintenance costs of these schemes are usually very high. The current estimation of YouTube's costs is 1 million dollars per day and these costs could increase extremely if more videos continue to be switched to greater qualities [60]. The high bandwidth required by live streaming video greatly limits the number of clients that can be served by a source. In fact, many streaming services today offer relatively low resolution in order to save bandwidth. The quality of those streaming services is typically not comparable to that in traditional TV networks. Resource management is thus a key issue in Internet streaming deployment. In client/server-based media streaming systems scenario, on the one hand, the processing power, storage capacity, and I/O throughput of the server may become the bottleneck; on the other hand, large number of long-distance network connections may also lead to traffic congestion. Hence, the system cannot meet the performance requirements of large-scale real-time media streaming applications [44].

Consider a situation in an on demand video service offered by Akamai to Doordarshan Online (http://www.dd.now.com). In April 2001, Doordarshan Online used Akamai's content distribution network to web-cast India-Australia cricket match. The company provisioned certain bandwidth from Akamai with its average number of clients in mind. As the match approached an exciting finish, the number of clients demanding the feed increased to surpass the provisioned bandwidth. The servers went down leading to disrepute of the site and annoyance among end-users. The above example shows that unicast schemes scale badly for flash crowds. In spite of growing the server resources, the surge in number of requests leads to saturation of server resources. Current trends indicate that such problems will exacerbate in the near future as the potential for demand intensifies, more events will be web-cast live by various companies to feed an increasing client base. It is rational to anticipate that as the edge bandwidth increases; the size of flash crowds will also increase, corresponding to taller spikes in traffic [43]. However, the IP multicast technique being offered to address these problems need support from special hardware and the costs of infrastructure setup and administration are expensive. In essence, the traditional techniques cannot resolve the problems of video streaming effectively [44]. Hence, an alternate mechanism is required to prevail over the resource saturation.

The emerging distributed information sharing architecture, P2P networks has been widely accepted as a means, to address the resource problem with Internet streaming applications like VoD and live streaming and to provide an alternative for client/server computing. Still in its infancy, both live and on-demand P2P streaming have the potential of altering the means we watch TV, providing ubiquitous access to an enormous number of channels.

## Peer-to-Peer Networks

The World Wide Web (WWW) can be viewed as a massive distributed system consisting of millions of clients and servers for accessing associated documents. Servers preserve collections of objects, whereas clients provide users a user-friendly interface for presenting and accessing these objects. The inadequacy of the client-server model is evident in WWW. Being resources are concentrated on one or a small number of nodes and to provide 24/7 access with satisfactory response times, complicated load-balancing and fault-tolerance algorithms have to be employed. The same holds right for network bandwidth, which adds to this tailback situation. These two key problems inspired researchers to come up with schemes for allocating processing load and network bandwidth among all nodes participating in a distributed information system [45].

P2P networks are a recent addition to the already large number of distributed system models. P2P networking has spawned immense attention worldwide amongst both Internet users and computer professionals. P2P computing takes advantage of existing computing power, computer storage and networking connectivity, allowing users to leverage their collective power to the 'benefit' of all. P2P system is defined as '*a self organizing system of equal, autonomous entities (peers) which aims for the shared usage of distributed resources in networked environment avoiding central services'*. Nodes in a P2P network usually play equal roles, so these nodes are also called peers. In this chapter the terms 'peer' and 'node' are used interchangeably in the context of P2P networks. The peers cooperate in a distributed manner to achieve the desired objective. The most important characteristics of P2P technology are direct interaction and data exchange between peer systems rather than through a central server. This is the basis for decentralized distributed computing. P2P networks are self-organized and adaptive. Peers may come and go freely. P2P systems handle these events automatically [46].

One of the main features in a P2P system is that each node contributes resources including bandwidth, storage space, and CPU power, and consequently, the entire system capacity can in fact increases as more nodes enter the a system. This is piercingly contrary to the client and server architecture, in which the addition of clients always degrades the overall performance. Another benefit that P2P brings is the robustness in case of failure as each node does not rely on any centralized server for content retrieval. In a P2P system, participating nodes mark at least part of their resources as 'shared', allowing other contributing peers to access these resources. Thus, if node A publishes something and node B downloads it, then when node C asks for the same information, it can access it from either node A or node B. As a result, as new users access a particular file, the system's capability to provide that file increases [47]. P2P networks have the prospective of diminishing the user perceived latency by pushing the data and computation to a location closer to the users.

Among the various P2P systems in different application domains, file sharing systems, where files are exchanged among peers, dominate the applications of P2P systems. P2P systems normally form, at the application level, a decentralized overlay network with its own routing mechanism. Until now, a few most important categories of P2P systems have been introduced with their own merits and demerits. Generally P2P systems are categorized as *centralized, decentralized structured* and *decentralized unstructured* P2P systems. In the beginning, a P2P system started out with a centralized index system where file locations are indexed in a number of selected servers for speedy searches. In the centralized model, such as Napster [48], central index servers are used to maintain a directory of shared files stored on peers with the intention that a peer can search for the location of a desired content from an 'index server'. Conversely, this design makes a single point failure and its centralized nature of the service generates systems vulnerable to denial of service attacks. The subsistence of a central authority introduced numerous legal problems such that the centralized index method was replaced by the decentralized index system. Subsequently, P2P systems have become completely decentralized in all their functions. Decentralized P2P systems have the advantages of eliminating reliance on central servers and providing freedom for participating users to exchange information and services directly between each other. Decentralized P2P systems can be categorized into two major systems: *unstructured* and *structured*.

In decentralized unstructured P2P systems, such as Gnutella there is neither a centralized index nor any strict control over the network topology or file placement. By and large, the peers self-configure into an overlay network with no particular intended topology. Distribution of files is probably managed in an ad-hoc way not designed to result in any particular arrangement. Since these systems have no coupling between the network topology and data placement, locating a desired file is not simple. Nodes joining the network, following some loose rules, form the network. In these systems, data are stored anywhere in the system and searched for by broadcasting queries to all peers within a specified distance. These methods are simple, and highly robust to alteration in the overlay network topology. Conversely, the ineffectiveness of broadcasting raises doubts about their scalability.

In decentralized structured models, such as Chord [49], Pastry [50], and CAN [51], the shared data placement and topology characteristics of the network are robustly controlled on the basis of distributed hash functions. The index is distributed in a precise way across the overlay network topology. The result is that queries are directed resourcefully towards the exact index location, solving the scalability problem of unstructured methods. On the other hand, structured schemes have troubles of their own: complexity, high maintenance overhead, a rigid structure that is somewhat fault intolerant and inability to support range and keyword queries, which are quite popular in P2P applications. Due to these drawbacks, structured methods have not so far been deployed on any broad scale [52].

Until recently, Internet P2P systems assumed all peers are identical and uniform in resources. Functionality is thus distributed without considering real-world heterogeneity of peer capabilities. For example, some peers may have smaller disk and slower processor speed than others. But they perform the same role and responsibility as other peers with greater capabilities. This results in instances of inefficiency and bottlenecks in performance due to very limited capabilities of these peers. To account for and even exploit the existence of such heterogeneity of peer capabilities, the notion of super-peers, which are well-provisioned in terms of resource capacity, have recently been introduced. A super-peer often plays the role of a server that manages the queries and responses for a subset of ordinary peers [53]. A super-peer acts as a server to a set of clients in the system and the whole set of super-peers are regarded as a centralized server to the clients just like Napster. So, this approach basically forms a hierarchical overlay network, where the top layer contains the super-peers, and the bottom layer consists of the peers. KaZaA is a P2P file-sharing application, which employs the idea of 'superpeers'. The notion of superpeers has been proposed in a recent version of the Gnutella protocol to perk up the scalability of its original system.

*Challenges in P2P Streaming*

Over the past few years, P2P networks have appeared as an auspicious method for the delivery of multimedia content over a large network. The intrinsic characteristics make the P2P model a potential candidate to solve various problems in multimedia streaming over the Internet [54]. The P2P streaming is more elegant because of two reasons. First, P2P does not need support from Internet routers and thereby cost effective and simple to deploy. Second, a peer simultaneously acts as a client as well as server, thus downloading a video stream and at the same time uploading it to other peers watching the program. Consequently, the P2P streaming significantly decreases the bandwidth needs of the source [58]. The objective of P2P streaming mechanisms is to maximize delivered quality to individual peers in a scalable fashion in spite of the heterogeneity and irregularity of their access link bandwidth. The aggregate available resources in this approach physically grow with the user population and can potentially scale to any number of participating peers [55]. Each peer should continuously be able to offer suitable content to its connected peers in the overlay by making use of outgoing bandwidth of participating peers [56]. However, providing P2P video streaming services for a large number of viewers creates very difficult technology challenges on both system and networking resources.

While traditional P2P file distribution applications target flexible data transfers, P2P streaming focuses on the efficient delivery of audio and video content under stiff timing requirements. Stream data are instantaneously received, played, and passed to other associated peers. For example, the P2P file sharing application - BitTorrent permits peers to interchange any segment of the content being distributed since the order in which they arrive is not important. In contrast, such techniques are not viable in streaming applications [57]. Video files are directly played-out while they are being downloaded. Therefore, pieces, which are received after their play-out time, degrade user experience. This degradation is visible either as missing frames or as a playback stop, which is also denoted by stalling. While redundancy schemes might be suitable for streaming because they do not require further communication between sender and receiver, retransmission might not be possible, because of the strict timing requirements. In addition, peers have limited upload capacities, which stems from the fact that the Internet was designed for the client/server paradigm and applications. Furthermore, the streaming systems suffer from packet drop or delay due to network congestions [60].

In a P2P streaming, the end-to-end delay from the source to a receiver may be excessive because the content may have to go through a number of intermediate receivers. The behavior of receivers is unpredictable; they are free to join and leave the service at any time, thus discarding their successor peers. Receivers may have to store some local data structures and exchange state information with each other to preserve the connectivity and to perk up the effectiveness of the P2P network. The control overhead at each receiver for satisfying such purposes should be small to keep away from excessive use of network resources and to overcome the resource limitation at each receiver. This is important to the scalability of a system with a large number of receivers [59].

Organizing the peers into a high quality overlay for disseminating the video stream is a challenging problem for broadcasting video in P2P networks. The constructed overlay must be effective both from the network and the application outlooks as broadcasting video concurrently requires high bandwidth and low latencies. On the other hand, a start-up delay of a couple of seconds is abided for applications which are real-time. The system should be able to accommodate tens of thousands of receivers at a time. At the same time, the overheads associated must be reasonable even at large scales. The construction of overlay must take place in a distributed fashion and must be robust to dynamic changes in the network. The system must be self-improving in that the overlay should incrementally progress into a better structure as more information becomes available [58].

*Approaches for Overlay Construction*

Existing streaming techniques in the P2P approach can be categorized into schemes supporting *P2P live video streaming* and those that support *P2P on demand video streaming*. Some techniques can offer both services. Several P2P streaming systems of above two categories have been deployed to provide on demand or live video streaming services over the Internet.

Based on the overlay network structure, P2P streaming systems are broadly classified into three categories: *tree-based, mesh-based* and *hybrid* schemes (Figure 6). The tree-based approaches use push based content delivery; however the mesh-based approaches use swarming content delivery. Several P2P live streaming and video on demand applications are built on these schemes. This section briefly discusses all the three categories of overlays along with example applications.

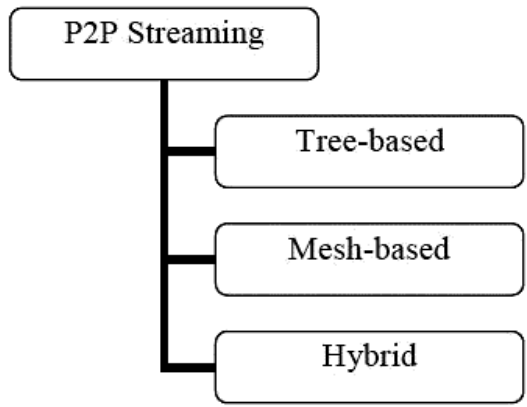

**Figure 6: P2P Streaming Types**

***Tree Based Overlay:*** Similar to an IP multicast tree formed by routers at the network level, users participating in a video streaming session can form a tree at the application layer that is rooted at the video source server. Tree-based overlays implement a tree distribution graph, rooted at the source of content (figure 7). In principle, each node receives data from a parent node, which may be the source or a peer. The tree-based systems typically distribute video by actively pushing data from a peer to its children peers [62].

A common approach to P2P streaming is to organize participating peers into a *single tree-structured overlay* over which the content is pushed from the source towards all peers e.g. – [61]. This way organizing peers is called *single-tree streaming*. In these systems, peers are hierarchically organized in a tree structure where the root is the stream source. The content is spread as a continuous flow of information from the source down to the tree. Each user joins the tree at certain level. All the load is supported by the interior nodes of the tree while leafs are just receiving data. Systems belonging to this category mainly differ in the algorithms used to create, and maintain the tree structure. Given a set of peers, there are many possible ways to construct a streaming tree to connect them up. The goal of tree construction algorithm is to maximize the bandwidth to the root of all nodes. Since these systems are very close to IP multicast, trying to emulate its tree structure, they are able to achieve data paths that do not differ too much from IP multicast paths.

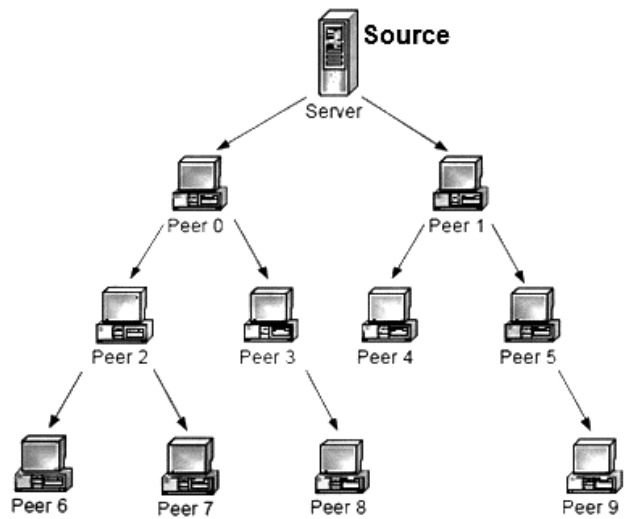

**Figure 7: Single Tree Model [58]**

Tree construction and maintenance can be done in either a *centralized or a distributed fashion* in single-tree streaming systems. In a centralized solution (Figure 8 & 9), a central server controls the tree construction and recovery. When a peer joins the system, it contacts the central server. Based on the existing topology and the characteristics of the newly joined peer, such as its location and network access, the server decides the position of the new peer in the tree and notifies it which parent peer to connect to. The central server can detect a peer departure through either a graceful sign-off signal or some type of time-out based inference. In both cases, the server recalculates the tree topology for the remaining peers and instructs them to form the new topology [65]. For a large streaming system, the central server might become the performance bottleneck and the single point of failure. To address this, various distributed algorithms, e.g. ZigZag [63], have been developed to construct and maintain streaming tree in a distributed way. If peers do not change too often, single tree based systems require little overhead as packets are forwarded from peer to peer without the necessity of additional messages. However, in high churn environments, the tree would be frequently damaged and reconstructed. This process requires considerable control message overhead. Consequently, peers must buffer data for at least the time required to repair the tree, in order to evade packet loss [62].

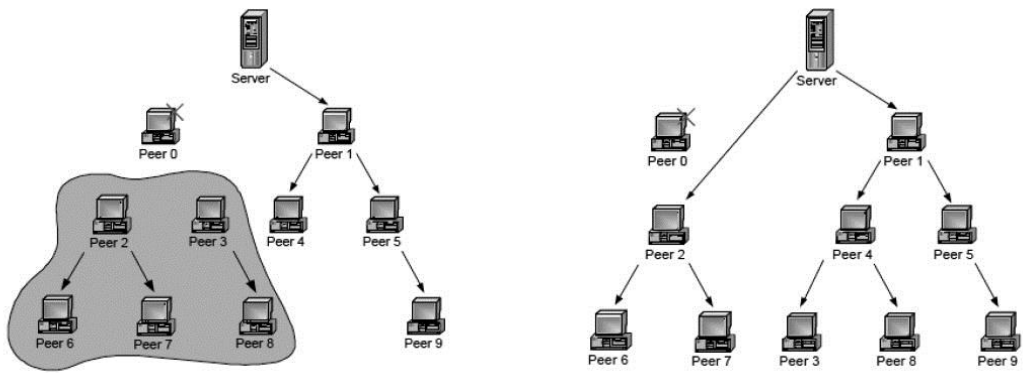

**Figure 8: Streaming Tree Reconstruction (a) Peer 0 departs (b) Tree Recovery after Churn [58]**

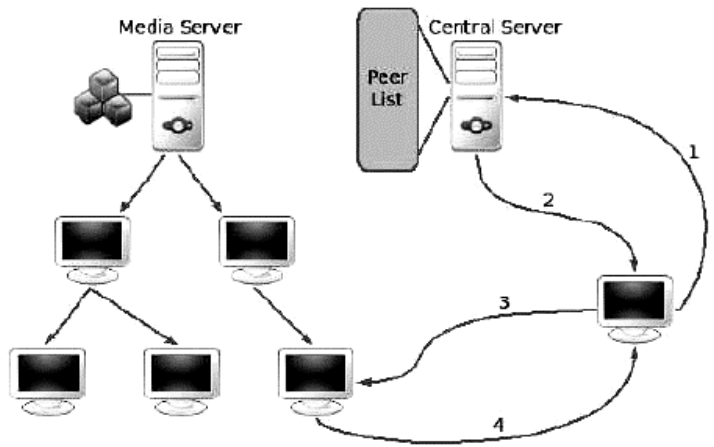

**Figure 9: Centralised Solution for Tree construction and Maintenance**
The peer sends its join request to the central server (1)
The central server sends proper providers for the peer and informs the peer about them (2)
The peer contacts them (3), and then the provider(s) sends data to peer (4).

*Tree Based Live Streaming Systems:* The most popular system using a single tree approach is **NICE** [64]. NICE is an acronym that stands for "*NICE the Internet Cooperative Environment*". NICE was initially designed for low-bandwidth, data streaming applications with a large number of receivers. The protocol arranges the set of end hosts into a hierarchy based on round-trip-time information between hosts (Figure 10). The basic operation of the protocol is to create and maintain the hierarchy. The hierarchy implies the routes. Logically, each member keeps detailed state about other members that are near in the hierarchy, and only has limited knowledge about other members in the group. The hierarchical structure is also important for localizing the effect of member failures. While constructing the NICE hierarchy, members that are "close" with respect to the distance metric are mapped to the same part of the hierarchy: this produces trees with low stretch.

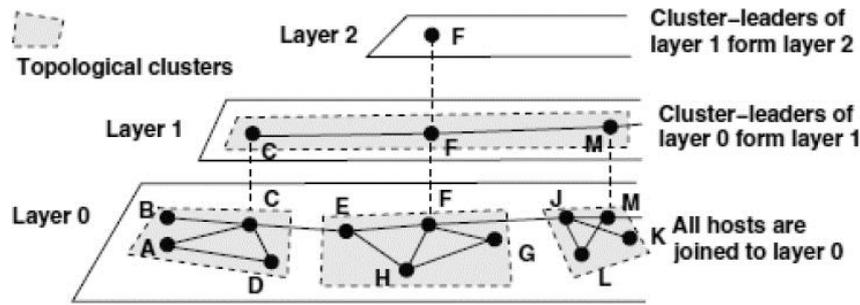

Figure 10: Hierarchical Arrangement of Hosts in NICE [64]

***SpreadIt*** [43] builds an application level multicast tree over the set of clients. Nodes are organized into different levels (Figure 11 & 12). For each node, n, at level l+1; l=0; 1; 2:::, there is a node, called its parent, p, at level l; n is called a child of p. All nodes in the sub-tree rooted at p are called its descendants. Each peer within the tree is responsible for forwarding the data to its descendants. Each client node needs to be enabled with a basic peering layer between the application and transport layers. Peering layers at different nodes coordinate among themselves to establish and maintain a multicast tree. The application (RealPlayer, Windows Media Player, etc.) gets the stream from the peering layer on their local machines. SpreadIt uses only a single distribution tree and hence is vulnerable to disruptions due to node departures.

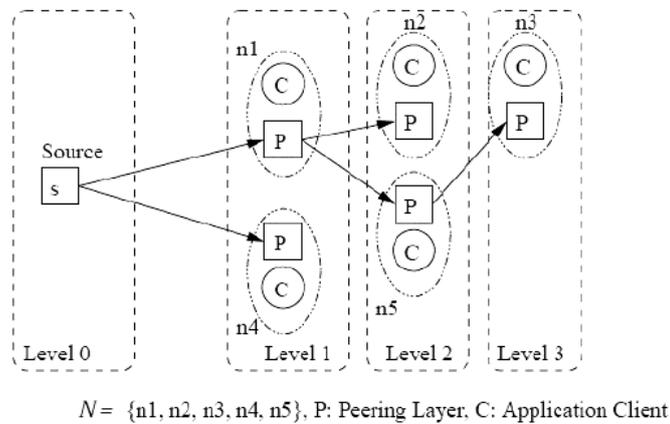

Figure 11: SpreadIt - An Application Level Multicast Tree Built on the Peers [50]

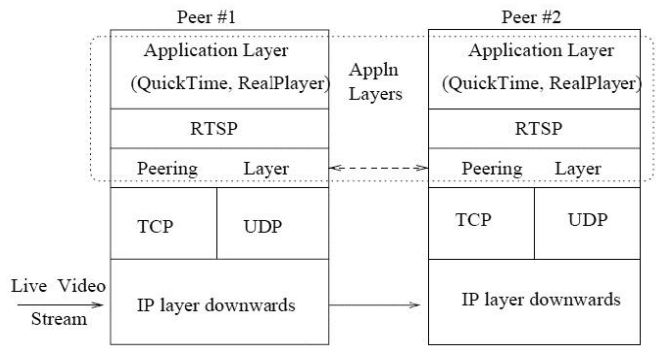

Figure 12: SpreadIt - A Layered Architecture of a Peer [43]

***End System Multicast (ESM)*** [66] is an infrastructure for media broadcasting, implemented by Carnegie Mellon University. ESM allows broadcasting audio/video data to a large pool of users. The ESM system employs a structure-based overlay protocol which constructs a tree rooted at the source. The information is delivered following a traditional single-tree approach, which implies that any given peer receives streams from only one source. Each ESM node maintains information about a small random subset of members, as well as information about the path from the source to itself. A new node joins the broadcast by contacting the source and retrieving a random list of members that are currently in the group. It then selects one of these members as its parent using the parent selection algorithm. To learn about members, a gossip-like protocol is used. Each node also maintains the application-level throughput it is receiving in a recent time window. If its performance is significantly below the source rate, then it selects a new parent as described in the parent selection algorithm. When a node joins the broadcast, or needs to make a parent change, it probes a random subset of nodes it knows. The probing is biased toward members that have not been probed or have low delay.

Figure 13 shows an example of ESM task. The end receivers could play the role of parent or children nodes. The parent nodes perform the membership and replication process. The children nodes are receivers who are getting data directly from the parent nodes. There is one central control server and one central data server residing in the same root source. Any receiver can play the role of parent to forward data to its children. Each client has two connections: *a control connection* and *a data connection*.

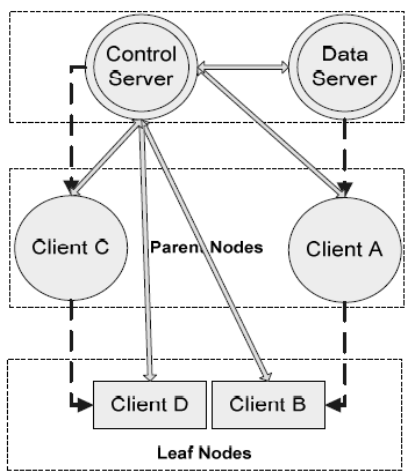

**Figure 13: Example of ESM**

One advantage of ESM is that it resolves the deployment problems of IP multicast. However, doing multicasting at end-hosts incurs in some performance penalties. Generally, end hosts do not handle routing information as routers do. In addition, the limitation in bandwidth and the need of forwarding messages from host-to-host using unicast connection, and consequently incrementing the end-to-end delay of the transmission process, contribute to the price to pay for this approach. These reasons make end-system multicast less efficient than IP multicast.

*ZigZag* [63] has been proposed by the University of Central Florida and that improves the NICE protocol. The tree organization is very close to the one proposed by NICE. The algorithms for structure building and maintenance are quite similar to NICE and all the NICE structure's properties are still valid. ZigZag organizes receivers into a hierarchy of clusters and builds the multicast tree atop this hierarchy according to a set of rules called C-rules (Figure 14). A cluster has a head and an associate head, the head responsible for monitoring the memberships of the cluster and the associate-head responsible for transmitting the content to cluster members. Therefore, the failure of the head does not affect the service continuity of other members, or in case the associate-head departs, the head is still working and can designate a new associate-head quickly. While in NICE everything is forwarded by the cluster leader, here the responsible for data forwarding is the associate head. The ZigZag protocol control overhead is low. A receiver needs to exchange control information to O(log N) other receivers in the worst case. On average, it communicates with at most a constant number of other receivers. ZigZag is best applicable to the streaming applications such as a single media server broadcasting a live long-term sport event to many clients, each staying in the system for a long enough period. As ZigZag is focused on the single-source media streaming, it is not suitable for the media streaming applications where multiple sources are presented. The main drawback of ZigZag is that it does not consider the upload bandwidth capacity of peers in join procedure. Also, because ZigZag creates single tree connection between peers, it has the general problems of single tree model, such as not using upload bandwidth of leaves and vulnerability to failure of interior nodes.

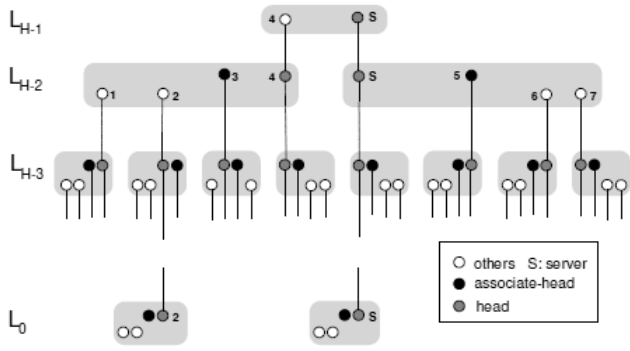

**Figure 14: Administrative Organisation of Peers in ZigZag [63]**

*Multi-tree based overlay:* Single tree-based solutions are perhaps the most natural approach, and do not require sophisticated video coding algorithms. However, one concern with single tree-based approaches is that the failure of nodes, particularly those higher in the tree may disrupt delivery of data to a large number of users, and potentially result in poor transient performance. If an interior node has not the required computational or bandwidth resources to serve all its children, peers in its sub-tree will suffer of high delays in data reception or will never receive the stream. The amount of data lost varies from one system to another and depends on the repairing mechanism being adopted. These systems don't seem to exploit very well all the available peers' resources and in particular the available bandwidth. For instance, the leaf nodes account for a large portion of peers in the system and they don't contribute their uploading bandwidth, which greatly degrades the peer bandwidth utilization efficiency. In response to these concerns, researchers have been investigating more resilient structures for data delivery. In particular, one approach that has gained popularity is *multi-tree based* approaches [58].

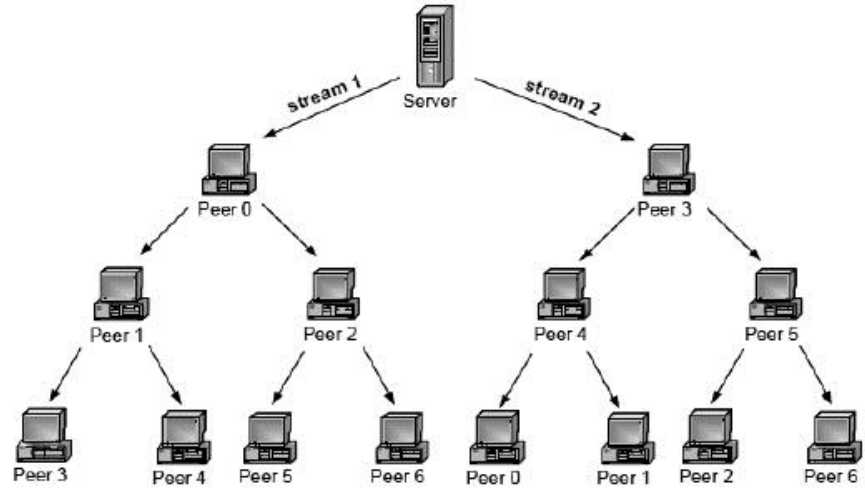

**Figure 15: Multiple Tree-based Streaming**

In the multi tree-based approach (figure 15), an overlay construction mechanism organizes participating peers into multiple trees. Each peer determines a proper number of trees to join based on its access link bandwidth. Each peer is placed as an internal node in only one tree and leaf node in other participating trees. When a peer joins the system, it contacts the bootstrapping node to identify a parent in the desired number of trees. In multiple-tree based P2P live streaming systems, the video is encoded into multiple sub-streams, and each sub-stream is delivered over one tree. The quality received by a peer depends on the number of sub-streams that it receives [65]. To keep the population of internal nodes balanced among different trees, a new node is added as an internal node to the tree that has the minimum number of internal nodes. To maintain short trees, a new internal node is placed as a child for the node with the lowest depth that can accommodate a new child or has a child that is a leaf. In the latter case, the new node replaces the leaf node and the partitioned leaf should rejoin the tree similar to a new leaf. When an internal node of a tree departs, each one of its child nodes as well as the subtree rooted at them are partitioned from the original tree, and thus should rejoin the tree. Peers in such a partitioned subtree initially wait for the root of the subtree to rejoin the tree as an internal node. If the root is unable to join the subtree after a certain period of time, individual peers in a partitioned subtree independently rejoin the tree with the same position as leaf or internal node. The content delivery is a simple push mechanism where internal nodes in each tree simply forward any received packets for the corresponding description to all of their child nodes. Therefore, the main component of the tree-based P2P streaming approach is the tree construction algorithm [67].

There are two key advantages for the multiple-tree solution. First, if a peer fails or leaves, all its children lose the sub-stream delivered from that peer, but they still receive the sub-streams delivered over the other trees. Due to this, all of its children would receive video streams in case of a loss of a sub-stream. Second, a peer plays different roles as internal node as well as leaf node in various trees. The upload bandwidth of an internal node can be utilized to upload the sub-stream delivered over that tree. At the same time, in order to provide high bandwidth utilization, a peer with a high upload bandwidth can supply sub-streams in several trees [65].

If peers do not change too often, multi-tree streaming systems require little overhead, since packets are forwarded from node to node without the need for extra messages. However, in high churn environments, the tree must be continuously destroyed and rebuilt. This process requires considerable control message overhead. Hence, nodes must buffer data for at least the time required to repair the tree, in order to avoid packet loss [68] [62].

*Multi-tree based Live Streaming Systems*

Few applications built-on multi-tree concept are available today. Examples are SplitStream and CoopNet.

**SplitStream** [69] is a multi-tree streaming system proposed in 2003 by the Microsoft Research center. The technique is designed to overcome the inherently unstable forwarding load in conventional tree-based multicast systems. The main idea of SplitStream is to split the stream into dissimilar independent stripes, and multicast each stripe using a separate tree. To ensure that the forwarding load can be spread across all participating peers, a forest of stripe trees is constructed in a way that a node is an interior node in at most one stripe tree and is a leaf node in all the other ones. Such a set of trees is called *interior-node-disjoint*. Figure 16 illustrates how SplitStream balances the forwarding load among the participating peers. In this example, the original content is split into two stripes and multicast in separate trees. Each peer, other than the source, receives both stripes. Each peer is an internal node in only one tree and forwards the stripe to two children. When an overloaded node receives a request from a prospective child, it either rejects this child or accepts it and rejects one of its existing children, which is less desirable than the new child. A node is more desirable if its node id is closer to its parent node id. In both cases the rejected child contacts one of the children of the overloaded node.

SplitStream builds the multicast trees for the stripes while respecting the inbound and outbound bandwidth constraints of the peers. It offers resilience to node failures and unannounced departures, even while the affected multicast tree is repaired. One of the main problems with SplitStream is the impact of nodes with heterogeneous bandwidth on its efficiency. Another problem is that in an interior-node-disjoint, nodes receive distinct stripes with different latencies as nodes are decisively placed in different distances from the root of multiple trees. This is undesirable for a live media streaming application, which involves strict timing constraints. The problem is augmented when the system scales to trees with larger depth and nodes are placed in diverse distances from the source. The former will either increase the source-to-end delay or disrupt the continuity of the media; while the latter wastes the bandwidth of both sender and receiver and unnecessarily burdens the network.

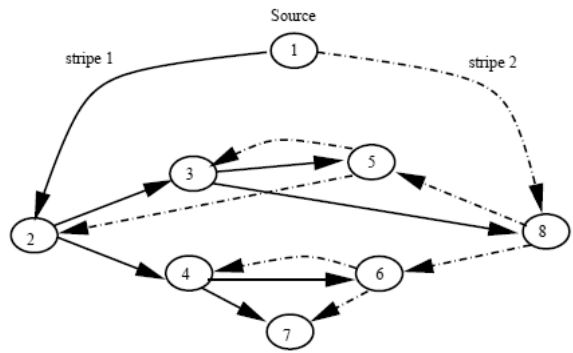

**Figure 16: A simple example illustrating the basic approach of SplitStream [69]**
The original content is split into two stripes. An independent multicast tree is constructed for each stripe such that a peer is an interior node in one tree and a leaf in the other.

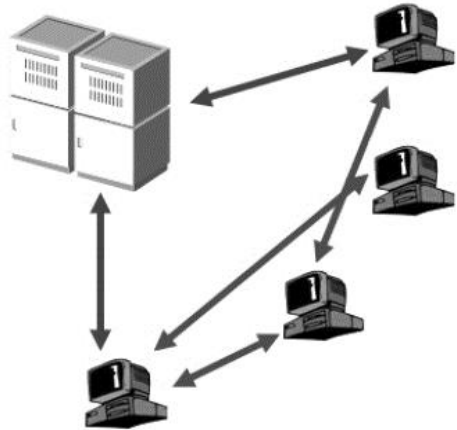

**Figure 17: Streaming Media Content Using CoopNet**

**CoopNet** (Cooperative Networking) [70] combines aspects of infrastructure-based and peer-to-peer content distribution. It adopts multiple description coding to carry on media data layering treatment, and then transmits media data in different layers along different tree paths. A resourceful server plays a central role in constructing and managing the distribution trees, whiles the bandwidth for forwarding the media data stream is still contributed by the distributed set of peers (Figure 17). The system builds multiple distribution trees spanning the source and all the receivers. When a node wants to join, it contacts the central server, which responds with a designated parent node in each tree. When a node leaves gracefully, it informs the central server, which will find a new parent for the children of the departed node and notifies the children of the identities of their new parent. CoopNet supports both the live streaming as well as on-demand streaming services. The CoopNet approach is good where a lower quality content presentation is preferred over loss of or delayed quality content presentation. However, since the central server needs to maintain full knowledge of all the distribution trees, it will put a heavy control overhead on the server. So, the scalability is not very good. Another problem is that the central server does constitute a single point of failure (Okuda, 2006).

*Mesh-based overlays:* To combat the peer dynamics, many recent P2P streaming systems use mesh-based streaming approach. In mesh-based approach, participating peers form a randomly connected overlay, or a mesh. In these overlays the original media content from a source is distributed among different peers. Each node knows every other node in the system. As a result, each node maintains connections with quite a few other nodes (neighbors) in the network. Each peer exchanges the data with a set of neighbors. If one neighbor leaves, the peer can still download the video from the remaining neighbors. Meanwhile, the peers will add other peers into its neighbor set. Unlike single tree systems in mesh-based systems, each peer can receive data from multiple supplying peers. Thus, mesh-based streaming systems are robust against peer dynamics. The major challenges in mesh-based P2P live streaming systems are neighborhood formation and data scheduling [65].

Upon arrival, a peer in the network contacts a bootstrapping node (tracker) to receive a set of peers that can potentially serve as parents. This approach is very similar to BitTorrent (Figure 18). The main advantage of this swarming content delivery is the ability to effectively utilize the outgoing bandwidth of participating peers as the group size grows [67]. The tracker provides a list of peers containing the information of a random subset of active peers available. Using this list, the peer attempts to initiate peering connections; and if successful, it starts exchanging video content with its neighbors. To handle unexpected peer departures, peers regularly exchange keep-alive messages. At the same time, depending upon system's peering strategies, a peer does not only connect to new neighbors in response to peer departures, but also when better streaming performance can be achieved.

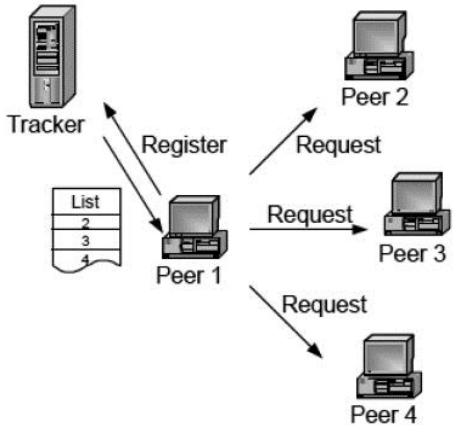

**Figure18: Peer List Retrieval from Tracker Server [71]**

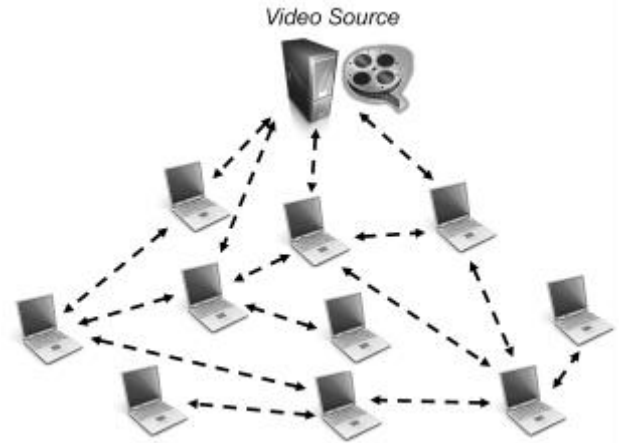

**Figure 19: P2P Live Video Streaming**

In mesh-based systems, the concept of video stream becomes invalid due to the mesh topology. The basic data unit in mesh-based systems is a video chunk. The multimedia server divides the media content into small media chunks of a small time interval, each of them with a unique sequence number that serves as a sequence identifier. Later, each chunk is disseminated to all peers through the mesh (Figure 19). Since chunks may take different paths in order to reach a peer, they may arrive to destination in a non-sequential order. To deal with this matter, received chunks are normally buffered into memory and sequentially rearranged before delivering them to its media player, ensuring continuous playback [65].

Mainly there are three major flavors of data exchange designs in mesh-based systems: *push, pull and hybrid push-pull* (Figure 20). In a *mesh-push* system, a peer actively pushes a received chunk to its neighbors who have not obtained the chunk yet. There is no clearly defined parent-child relationship in mesh-based system. A peer might blindly push a chunk to a peer already having the chunk. It might also happen that two peers push the same chunk to the same peer. Peer uploading bandwidth will be wasted in redundant pushes. To address that problem, chunk push schedules need to be carefully planned between neighbors. And the schedules need to be reconstructed upon neighbor arrivals and departures.

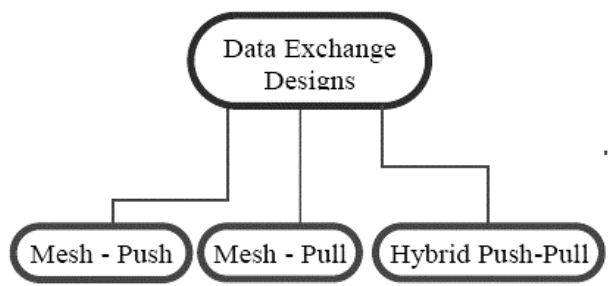

**Figure 20: Data Exchange Designs in Mesh-based Systems**

Another method for data delivery is *pull* method. The main idea of pull method is that each peer explicitly requests required chunk from other peers. Each peer has a neighbor set and it periodically exchanges data availability information (buffer maps) with its neighbors. A buffer map contains the sequence numbers of chunks currently available in a peer's buffer. Whenever a peer receives such information from other peers, it learns about the chunks it has not received yet. It then requests the missing chunks from the peers in the neighbor set, who possess it. Redundancy is avoided, as the node pulls data only if it does not already possess it. Further, since any chunk may be available at multiple partners, the overlay is robust to failures-departure of a node simply means its partners will use other partners to receive data segments. Finally, the randomized partnerships imply that the potential bandwidth available between the peers can be fully utilized [71]. A disadvantage of the pull technique is that both frequent buffer map exchanges and pull requests produce more signaling overhead and introduce additional delays while retrieving a chunk.

The pull mode in the unstructured overlay which is inherently robust can work well with the high churn rate in P2P environment while the push mode can efficiently reduce the accumulated latency observed at user nodes. The pure pull method can't meet the demands of delay-sensitive applications because of the striking latency accumulated hop by hop. Additionally, strong buffer capacities at each node are needed to store the exchanging data. The *hybrid push-pull* [72] streaming can greatly reduce the latency and inherit most good features such as simplicity and robustness of the pure pull method. Each node uses the pull method as a startup, and after that each node will relay a chunk to its neighbors as soon as the packet arrives without explicit requests from the neighbors. The streaming packets are classified as *pulling packets* and *pushing packets*. A pulling packet of a node is delivered by a neighbor only when the packet is requested, while a pushing packet is relayed by a neighbor as soon as it is received. Each node works under pure pull mode in the first time interval when just joining. After that, based on the traffic from each neighbor, the node will subscribe the pushing packets from its neighbors accordingly at the end of each time interval. A simple roulette wheel selection scheme is employed to allocate pushing packets in the next time interval to each neighbor. The selection probability of a neighbor is equal to the percentage of traffic from that neighbor in the previous time interval. Meanwhile, the lost packets induced by the unreliability of the network link or the neighbors failure will be pulled as well from the neighbors, where the roulette wheel selection scheme is also used to select the suppliers of each packet from neighbors. Thus, most of the packets received will be pushing packets from the second time interval.

*Popular Mesh-based Live Streaming Systems*

Several applications are developed by researchers for various categories of mesh based P2P streaming. AnySee is a push based streaming application whereas CoolStreaming, Chainsaw, PPLive, PPStream, and SopCast are examples of pull based streaming applications. GridMedia and PRIME are applications developed based on hybrid scheme called push-pull approach.

**AnySee** [73] is a mesh push based streaming system in which resources are assigned based on their locality and delay. The basic workflow of AnySee is as follows. Initially, a mesh-based overlay is constructed. Every peer, with a unique identifier, first connects the bootstrapping peers and selects one or several peers to construct logical links. Each peer thus maintains a group of logical neighbors. A location detector based algorithm is employed to match the overlay with the underlying physical topology. Initially, all streaming paths are managed by the *single overlay manager* which deals with the join/leave operations of peers. The *inter-overlay optimization manager* explores appropriate paths, builds backup links, and cuts off paths with low QoS for each end peer. The manager maintains two sets of active streaming paths, including the current streaming path and the pre-computed backup paths, of all the peers in the network. So, when a peer fails or leaves the network selfishly, a new path is selected from the backup sets to replace the broken link thus restoring the connectivity of the network. The system diagram of an AnySee node is shown in figure 21. This mechanism is advantageous because the neighboring peers are not swarmed with requests due to a peer's departure; instead the overlay manager just replaces a lost link by referring to the backup set and thus replacing a peer efficiently. Hence, AnySee restores the connectivity of the network very quickly [74] since the restoration plan for the descendant peers of the missing peer is carried out beforehand. The *key node manager* allocates the limited resources, and the *buffer manager* manages and schedules the transmission of media data. The goal of the key node manager is to determine the number of requests that a peer should have. Videos are partitioned into chunks, each with a fixed playing time of 1s. Peers fetch chunks from sources or peers and cache them in local memory. The weakness of AnySee is that the media quality cannot be guaranteed, since a group of randomly selected peers may not have enough resources to provide the desired media quality.

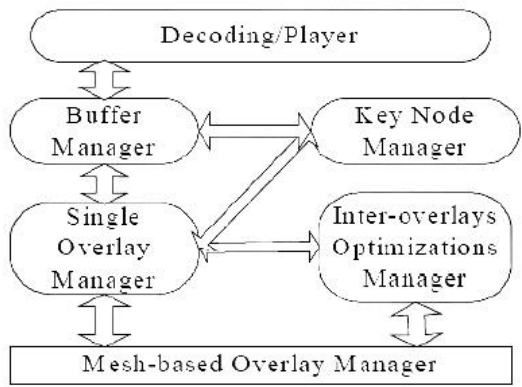

**Figure 21: The system diagram of an AnySee node [73]**

**Chainsaw** [75] is a mesh pull-based system that does not rely on a rigid network structure. In this scheme, peers are notified of new packets by their neighbors and must explicitly request a packet from a neighbor in order to receive it. This way, duplicate data can be eliminated and a peer can ensure it receives all packets. Every peer maintains a *window of interest*, which is the range of sequence numbers that the peer is interested in acquiring at the current time. It also maintains and informs its neighbors about a *window of availability*, which is the range of packets that it is willing to upload to its neighbors. The window of availability will typically be larger than the window of interest. For every neighbor, a peer creates a list of *desired packets*, i.e. a list of packets that the peer wants, and is in the neighbor's window of availability. It will then apply some strategy to pick one or more packets from the list and request them via a request message. A peer keeps track of what packets it has requested from every neighbor and ensures that it does not request the same packet from multiple neighbors. It also limits the number of outstanding requests with a given neighbor, to ensure that requests are spread out over all neighbors. Nodes keep track of requests from their neighbors and send the corresponding packets as bandwidth allows. The system does not provide a mechanism to enforce a fair resource contribution as Chainsaw allows peers to define its own maximum uploading bandwidth and fails to deter free riding [76]. Chainsaw can potentially invite high network and CPU overheads due to per packet announcements.

**PPLive** is a mesh pull P2P streaming platform that distributes both live and pre-recorded contents. The major difference with BitTorrent is that in PPLive packets must meet the playback deadline. In January 2008 the PPLive application provided almost 500 channels with 1,000,000 daily users on average. The number of channels in December 2008 was reported to be equal to 1775. The PPLive platform consists of multiple overlays. A single overlay corresponds to a PPLive channel. Each peer in an overlay is identified by the pair (IP address, port number). *Figure 22* shows the basic actions of a PPLive peer. At first, PPLive peer downloads channel list from the *channel list server* via http. After that for the selected channel, the peer collects a small set of peers involved in the same overlay by querying the *membership servers* via UDP. The peer communicates with the peers in the list to obtain additional lists which it aggregates with its existing peer list via UDP. In this manner the peer maintains a list of other peers watching the same channel [77]. In order to relax the time requirements, to have enough time to react to node failures and to smooth out the jitter, packets flow through two buffers: one is managed by PPLive, and the second by the media player. A downside of such architecture is the long start-up delay [78]. The working of PPStream is very similar to PPLive.

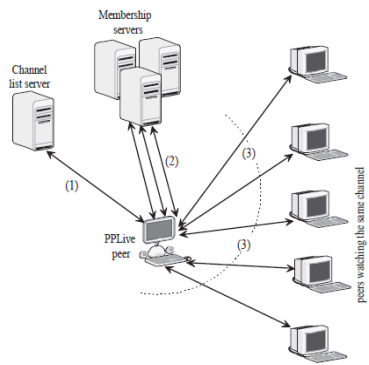

Figure 22: PPLive basic architecture

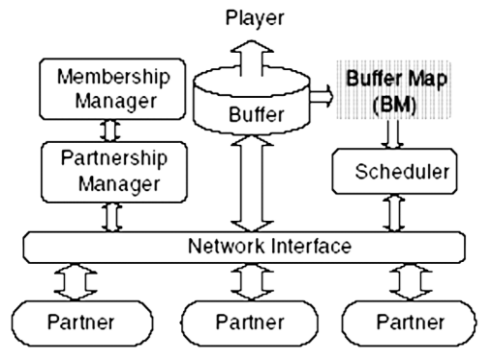

Figure 23: A generic system diagram for a DONet node [79]

**DONet** (or CoolStreaming) [79] is another successful mesh pull P2P streaming system implemented by the Universities of Hong-Kong and Vancouver. In DONet every node periodically exchanges data availability information with a set of partners, and retrieves unavailable data from one or more partners, or supplies available data to partners. A node consists of three key modules (figure 23): a *membership manager*, which helps the node maintain a partial view of other overlay nodes; a *partnership manager*, which establishes and maintains partnership with other nodes; *a scheduler*, which schedules the transmission of video data. The *scheduler* determines which segment should be obtained from which partner and downloads segments from partners and uploads their wanted segments. CoolStreaming requires newly joining nodes to contact the origin server to obtain an initial set of partner candidates. Each node also maintains a partial subset of other participants in the group. CoolStreaming employs Scalable Gossip Membership protocol (SGAM) to distribute membership messages. A CoolStreaming node can depart either gracefully or accidentally due to crash. In either case, the departure can be easily detected after an idle time and an affected node can quickly react through re-scheduling using the buffer map information of the remaining partners. CoolStreaming also let each node periodically establish new partnerships with nodes randomly selected from its local membership list. This operation helps each node maintain a stable number of partners in the presence of node departures and explore partners of better quality, for example those constantly having a higher upload bandwidth and more available segments [58]. CoolStreaming supported several different types of media players, such as Windows Media Player, Real Player or other media players. Using the scheduling algorithm and a strong buffering system, CoolStreaming achieves a smooth video playback and a very good scalability as well as performance. The overall streaming rate and playback continuity of CoolStreaming system is proportional to the amount of peers online at any given time [80]. One of the disadvantages of DONet is that notifying peers and afterward requesting segments possibly results in long delays before any data is exchanged. Similarly, due to the random selection algorithm, the quality of service cannot be assured. Moreover, DONet assumes that all the peers can cooperate in the replication of the stream; it is likely to have selfish peers in systems that do not want to share their upload bandwidth.

**SopCast** is a free BitTorrent-like P2PTV application, born as a student project at Fundan University in China. SoP is the abbreviation for Streaming over P2P. In SopCast the channels can be encoded in Windows Media Video (WMV), Video file for Realplayer (RMVB), Real Media (RM), Advanced Streaming Format (ASF), and MPEG Audio Stream

Layer III (MP3). A Client has multiple choices of TV channels, each of which forms its own overlay. Each channel streams either live audio-video feeds, or loop-displayed movies according to a preset schedule. The viewer tunes into a channel of his choice and SopCast starts its own operations to retrieve the stream. After some seconds a player pops up and the stream can be seen. It also allows a user to broadcast his own channel. SopCast provides low overall frame loss ratio. However, SopCast suffers from peer lags, i.e., peers watching the same channel might not be synchronized. Moreover, the zapping time is extremely high [81].

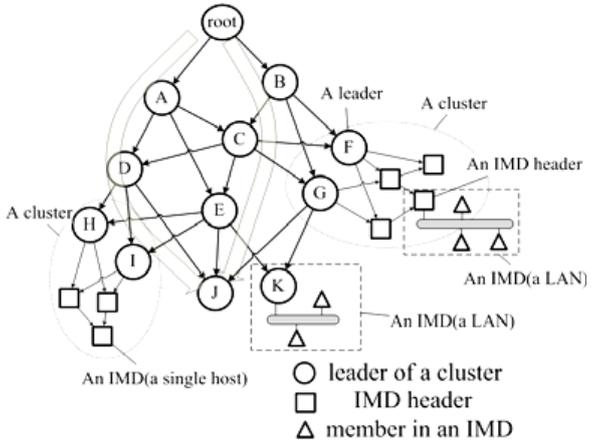

**Figure 24: GridMedia Architecture Based on MSOMP [82]**

**GridMedia** [82] adopts a push-pull streaming mechanism to fetch data from the partner nodes. The pull mode in the unstructured overlay can work well with the high churn rate in P2P environment while the push mode can reduce the accumulated latency at user side. A well-known rendezvous point (RP) - tracker server is deployed to assist the construction of the overlay. As a startup, a participating node first contacts the RP to get a list of part of the nodes already in the overlay, called a login process. Then the participating node will randomly select some nodes in this list as its neighbors. GridMedia mainly consists of *multi-sender based overlay multicast protocol (MSOMP)* and *multi-sender based redundancy retransmitting algorithm (MSRRA)*. MSOMP originates from the streaming server which is a node at the root. The MSOMP deploys mesh-based two-layer structure and groups all the peers into clusters with multiple distinct paths from the source root to each peer. Then with one or several leaders in each cluster, all the leaders construct the backbone of the overlay to build the upper layer. MSOMP provides each leader with multiple parents to receiver distinct streams simultaneously. MSOMP utilizes the existing IP multicast service which is available in many LANs. IP Multicast Domain (IMD) is a local network of any size that supports IP multicast. An IMD could be a single host, a LAN, etc. In each IMD, there is a header peer which is responsible for disseminating streaming content to other peers in the same IMD. As soon as the header leaves, a new header will be elected to replace the original role. MSOMP connects the IMDs by unicast tunnel altogether. MSOMP based GridMedia Architecture is shown in figure 24.

To address the problem of long burst packet loss, the MSRRA is proposed at the sender peers to patch the lost packets by using receiver peer loss pattern prediction. In MSRRA, each receiver peer obtains streaming packets simultaneously from multiple senders. Every sender peer transmits part of the streaming content. As soon as there is congestion occurring on one link, the receiver will take notice of this congestion and subsequently it notifies other senders who will continuously patch the episode of lost packets. The MSRRA algorithm efficiently relieves the impact of nodes failure, network congestion and link switch operations.

**PRIME** [83, 84] is a scalable push-pull mesh-based P2P streaming mechanism for live content. The foremost design goal of PRIME is to diminish bandwidth bottleneck and content bottleneck. PRIME incorporates swarming content delivery which combines *push* content reporting by parents with *pull* content requesting by children. Each peer simultaneously receives content from all of its parents and provides content to all of its children. Given the available packets at individual parents, a packet scheduling scheme at each peer periodically determines an ordered list of packets that should be requested from each parent. Parents simply deliver requested packets by each child in the provided order and at the rate that is determined by the congestion control mechanism. Each segment of the content is delivered to individual participating peers in two phases: *diffusion phase* and *swarming phase*. During the diffusion phase, each peer receives any piece of a new segment from its parent in the higher level. Therefore, pieces of a newly generated segment are progressively pulled by peers at different levels. During the swarming phase, each peer receives all the missing pieces of a segment from its parent in the same or lower levels. These parents are called *swarming parents*. Each piece of any new segment is diffused through a particular diffusion subtree during the diffusion phase of that segment. Then, the available pieces are exchanged between peers in different diffusion subtrees through the swarming mesh during the swarming phase of the segment. The application of the two different phases for content delivery leads to effective utilization of available resources to accommodate scalability and also minimizes content bottleneck. The disadvantage of PRIME is that if content bottleneck happens, nodes have to wait long in order to find their required data units after a few swarming phases, when that data becomes available in their neighborhood. Hence, there is no assurance for a reasonable level of streaming quality. Moreover, the algorithm doesn't consider the behavior of the P2P system in presence of a churn.

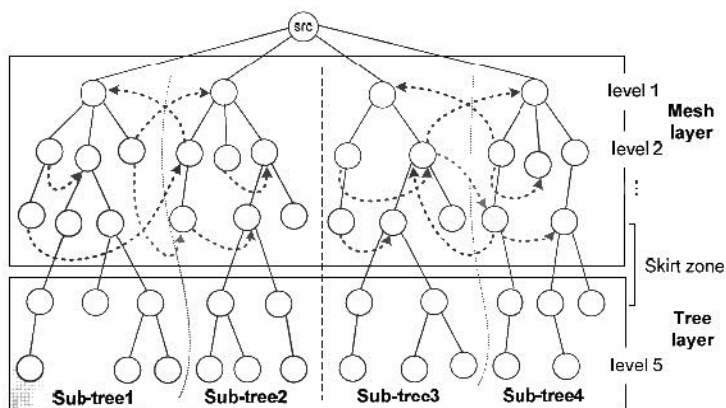

**Figure 25. Two-layer mesh/tree overlay in Hypo**

**HyPO** [85] is a hybrid P2P overlay for live media streaming. The scheme optimizes the overlay by organizing peers with similar bandwidth ranges in geographical area into a mesh overlay, and forms a tree overlay by selecting peers which are determined as stable. Figure 25 illustrates a two-layer mesh/tree overlay in Hypo. Depending on the tree optimization mechanism, the peers which have a large bandwidth will be near to the media source node in the tree overlay, and evenly distributed in the tree with branches of a similar depth. Consequently, the tree optimization reduces the average depth of the tree, thus enhancing the scalability. The mesh in HyPO is not an auxiliary connection since the peers in mesh member always delivers the data in mesh style until it becomes a tree member. However, since all procedures of HyPO rely on a bootstrap server, a discontinuous period may occur if the server fails. Furthermore, the HyPO does not mention how data are delivered in its mesh overlay [86].

*mTreebone* [87] is a collaborative tree-mesh design that leverages both mesh and tree structures. The key idea is to identify a set of stable nodes to construct a tree-based backbone, called *treebone*, with most of the data being pushed over this backbone. These stable nodes, together with others, are further organized through an auxiliary mesh overlay, which facilitates the treebone to accommodate node dynamics and fully exploit the available bandwidth between overlay nodes. Other non-stable nodes are attached to the backbone as outskirts. Figure 26 shows a mTreebone framework. In this scheme, the mesh connection is invoked only if there is an isolated node affected by parent departure or failure. The treebone maintenance and optimization only happen at the treebone nodes and there is no extra overhead for the outskirts peers. Normally, the streaming quality is much better for the treebone nodes due to the better stability of their data delivery paths from the source. The key challenge is that we need to identify the set of stable overlay nodes and position them at appropriate locations in the tree. Such a requirement can conflict with the bandwidth and delay optimization in tree construction. An additional complication when discussing stability is that this depends on human behavior - that is, on how long the user decides to stay [71]. Locality based clustering was not considered in mTreebone. On the other hand, CliqueStream [88], a hybrid overlay similar to mTreebone exploits the properties of a clustered P2P overlay to achieve the locality properties (Figure 27). CliqueStream elects one or more stable nodes of maximum available bandwidth in each cluster and allocates special relaying role to them. To maintain transmission efficiency, a content delivery tree is constructed out of the stable nodes using the structure in the underlying routing substrate and content is pushed through them. Less stable nodes within a given cluster then participate in the content dissemination and pull the content creating a mesh around the stable nodes.

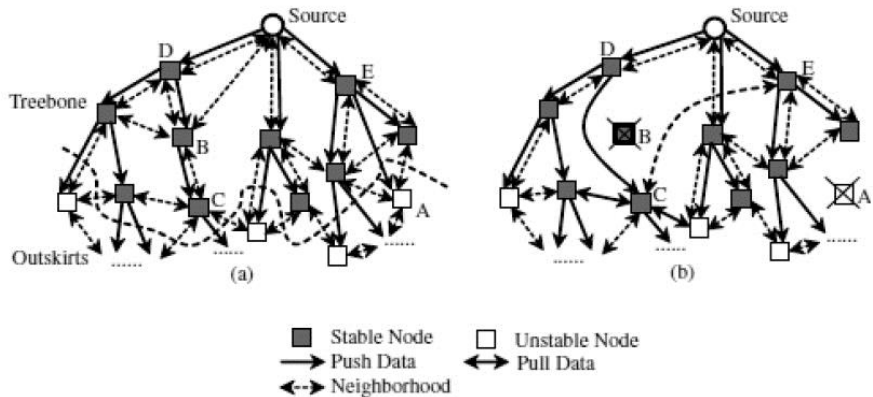

**Figure 26: mTreebone framework. (a) A hybrid overlay; (b) Handling node dynamics [87]**

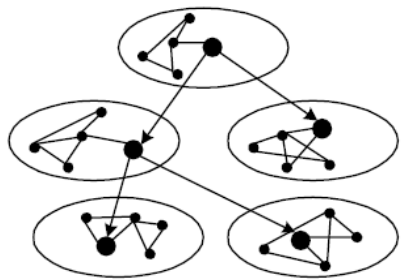

**Figure 27: Streaming topology in CliqueStream [88]**

The emerging hybrid push-pull P2P streaming overlays present a viable alternative for the traditional way of overlay construction such as tree and mesh since the hybrid design greatly simplifies the overlay construction and maintenance processes and at the same time largely retains its efficiency, and achieves fine-grained control over load.

## P2P on Demand Video Streaming

The existing VoD schemes poses several issues such as infeasibility of multicast, server crashes, and high maintenance and deployment costs of dedicated overlay routers. However, P2P based video streaming provides an alternative architecture for video on demand services. In a P2P VoD system, all peers are Internet-connected hosts, which store and stream the video to the requesting clients. The cost of these peers and Internet access would be bear by the clients rather than by providers of VoD service. Because there is an abundant supply of potential supplying peers with underutilized resources such as bandwidth and storage, P2P based architectures should have costs that are significantly less than the traditional client-server and CDN solutions [89].

Applying P2P live streaming techniques directly into VoD streaming is not a trivial task due to several reasons. Like P2P live streaming systems, the P2P-VoD systems also deliver the content by streaming. However, peers can watch different parts of a video at the same time, hence thinning their ability to help each other and relieve of the server [90]. A VoD capability would enable users to start watching a video after waiting for a small start-up time, while downloading the video in parallel. Even though, the shorter end-to-end delay makes live streaming more lively for the users, since in VoD streaming the video stream is previously recorded, the liveness is irrelevant. Hence, a short tree rooted at the video server and spanned over peers is not desirable in VoD streaming. The users should be able to watch the video at an arbitrary time, unlike in live streaming where they need to synchronize their viewing times. The users should also be able to perform control operations like rewind, forward etc... on the video [91]. Besides, the relationship between various variables is different for the two types of streaming. For example, a peer will likely stop watching a VoD stream when its QoS degrades, but the peer may not do the same thing for a live stream because he/she does not have an option of watching it again in the future. Therefore, it is expected that if the QoS of the video stream reduces, there will be many more peers leaving the system in VoD streaming case than the case of live streaming. This stretches the significance of a strong failure recovery protocol in a VoD streaming system. The protocol reconnects the abandoned peers efficiently, so that there are no loss of frame and no long delay at client's playback.

Another important requirement of a VoD service is *scalability*. A typical video stream imposes a heavy burden both on the network and the system resources such as disk I/O of the server. A VoD system should permit a new peer to join the system fast. The shorter the joining time is, the shorter the startup delay for a peer is. The joining requests of peers arrive to the system at different times. It is expected that the system must deliver the video in full-length to every peer without making the server become a bottleneck [92]. P2P-VoD systems usually require users to contribute larger amount of storage as these systems need huge buffer sizes in order to satisfy the diversified request from peers on different kinds of video programs. This storage space is usually 1 GB in PPLive. In effect after a user installs PPLive and run the system for the first time, the user could see an unknown kind of file of 1 GB exist in secondary storage [90].

Like video streaming systems, P2P VoD systems are generally classified as *tree based* and *mesh based* systems.

*Tree based VoD systems*

The users using tree-based overlay is synchronized and receive the content in the order the server sends it out. This is fundamentally different from the requirement imposed by VoD service. The major issue in tree based systems is the design of appropriate procedure for accommodating asynchronous users into the system. P2Cast and P2VoD are examples of tree based P2P VoD systems.

P2Cast [93] is an early patching scheme for VoD service. It is founded on the patching scheme proposed to support VoD service using native IP multicast. P2Cast addresses two key technical issues such as constructing an application overlay appropriate for streaming and providing a continuous stream playback in the face of disruption from an early departing client. The clients arriving within a threshold form a session. For each session, the server, together with the P2Cast clients, form an application-level multicast tree over the unicast-only network. The clients in P2Cast can forward the video stream to other clients, and also cache and serve the initial portions of a video to other clients. Every client actively contributes its bandwidth and storage space to the system while taking advantage of the resources located at other clients. The entire video is streamed over the application-level multicast tree, so that it can be shared among clients. For clients who arrive later than the first client in the session and thus miss an initial segment of the video, the segment can be retrieved from the server or other clients that have already cached that initial segment. P2Cast can serve many more clients than traditional client-server unicast service. The recovery scheme in P2Cast lets peers receive data from server directly when parent departure occurs. However, this increases the workload of the server.

P2VoD [94] is a tree based P2P video-on-demand scheme which tries to solve the problems of quick join, provides fast and localized failure recovery without jitter, effectively handles clients' asynchronous requests and provides small control overhead as compared to P2Cast. Each client in P2VoD has a *variable-size* FIFO buffer to cache the most recent content of the video stream it receives. Existing clients in P2VoD can forward the video stream to a new client as long as they have enough out-bound bandwidth and still hold the first block of the video file in the buffer. The failures are managed with the concept of *generation* and a *caching scheme*. The caching scheme allows a group of clients, arriving to the system at different times, to store the same video content in the prefix of their buffers. Such group forms a generation. When a member of a generation leaves the system, any remaining member of that generation can provide the video stream without jitter to the abandoned children of the leaving member provided that out-bound bandwidth is sufficient. In P2VoD, a streaming connection is assumed to be constant bit-rate, which equals to the playback rate of the video player. The recovery process in VoD is more complicated [95]. In addition, P2VoD does not consider the heterogeneous bandwidth of peers.

Cache and relay is a tree based approach in which a VoD client commonly relies on the content that resides in its parents' buffers. In this scheme, routers do not carry multicast functionalities. Hence, end-hosts are in charge for the caching and allocation of streaming media. The end-hosts may be client machines or proxies thereof and these systems maintain retrieved media objects in their local caches provisionally. If another client requests the media objects later on, the original server can forward the request to those end hosts who are physically closer to the client.

*oStream* [96] takes advantage of buffering capabilities of end hosts by employing cache and relay approach. The scheme employs a spanning tree algorithm for peers to construct an overlay for media streaming. oStream reduces the topological inefficiencies such as link stress and stretch introduced by using application layer multicast.

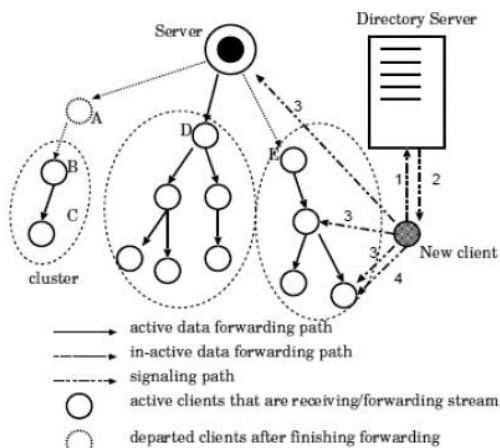

**Figure 28: DirectStream Architecture [97]**

A framework called *DirectStream* [97] allows clients to take advantage of the benefits of interval caching and video on-demand service with VCR operation support. DirectStream comprises a directory server, content servers, and clients. The directory server works as a central administrative point. It maintains a database that keeps track of all servers and clients participating in DirectStream, and helps new clients to locate the required service. The content servers provide the same functionality as in the traditional client-server service model storing contents in their repository and serving clients' requests so long as sufficient bandwidth is available. Thus the clients in DirectStream function as P2P nodes. A peer caches a moving window of the latest received content, and serves latecomers by continuously forwarding the cached content. A set of active clients among which a P2P streaming overlay is established is called a cluster. Clusters in DirectStream evolve over time and each client in a cluster share the same stream. The service search process for a new request consists of four steps, as indicated in Figure 28. First the new client sends a request to the directory server to ask for the video starting at position. The directory server then looks into its database and returns a list of candidate nodes, including both the content server and clients that have the content to serve this request. The new client determines from which node to retrieve the stream using the QoS parent selection algorithm. Using this algorithm, a client selects a parent node that has sufficient bandwidth. The new client contacts the selected candidate node and asks to forward the stream. After the connection is successfully set up, the new client signal back to the directory server and registers itself into the database. *DirectStream* significantly reduces the workload posed on the server. Another advantage is that it scales well as the popularity of the video increases even if participating clients behave non-cooperatively. DirectStream has two drawbacks. The centralized management presents a single point of failure. When numerous different ancestors fail, a peer can quickly starve its buffer.

In tree based VoD systems, peers on the upper layer always play an important role in the whole overlay network. Their departure will lead to the lower layer network fluctuation. Moreover, each peer has only one data supplier, which will cause inefficient utilization of available bandwidth in a heterogeneous and highly dynamic network environment. At the same time, in cache and relay based systems, if a parent jumps to another play point in the video, it starts to receive media data which is of no interest for its children, and those need to search for a new parent.

*Mesh based VoD systems*

In mesh based VoD systems, no specific topology is created. Peers in the network, based on the design rules, connect to several parents to receive video packets. Mesh-based VoD systems have lower protocol overhead, are much easier to design, are more resilient to high rates of churn, and hence are more popular. Current P2P mesh based systems have been shown to be very proficient for large-scale content distribution with few server resources. However, such systems have been designed for generic file distribution and provide a limited user experience for viewing media content. However, in VoD systems, the difficulty lies in the fact that users want to receive blocks "sequentially" in order to watch the movie while downloading. In addition, in VoD services, the users may be interested in different parts of the movie, and may compete for system resources. Over-all, the main challenge resides in designing systems that ensure that users can start watching a movie at any point in time, with small start-up times and sustainable playback rates [98].

*BitTorrent (BT)* is one of the most successful mesh P2P mechanisms for distributing huge volumes of content over the Internet. It is a scalable file sharing protocol which also incorporates swarming data transfer mechanism. There are several limitations of original BT strategy in providing video streaming. In BT, files are segmented on space. Although the default piece selection mechanism of BitTorrent is very efficient in minimizing the probability for rare pieces to become extinct and in providing peers with rare pieces, it fails despondently in case of time sensitive traffic. The reason is that with time sensitive data each piece must be received within a certain time limit. This factor is not taken into consideration in the original piece selection mechanism of BT and thus it cannot provide time sensitive distribution services, since pieces are requested based on their rareness and not by their deadline. Consequently, the current piece selection mechanism needs modifications in order to support a time-sensitive service such as VoD [99]. BASS and BiToS are examples of BitTorrent based mesh P2P VoD systems.

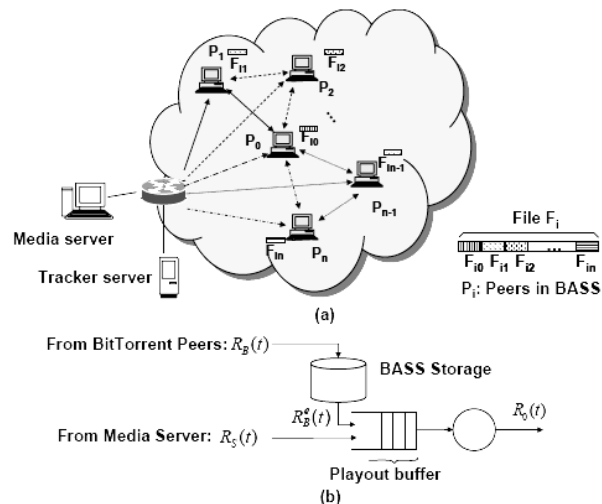

**Figure 29. BASS: (a) System Overview, (b) Client Model [100]**

*BitTorrent Assisted Streaming System (BASS)* [100] extends the current BitTorrent system to provide a near Video-on-Demand service. Since BASS uses the assistance of BT for streaming, it utilizes the service of an external server which stores all of the publisher's videos and guarantees that the users can playback the video at the playback rate without any quality degradation. The only modification to BitTorrent being that it should not download any data prior to current playback point. It is allowed to use rarest piece first and tit-for-tat policies. In rarest-piece-first policy, the client requests a piece based on the number of copies it sees available and choose the least common one. In tit-for-tat, a leecher (one who downloads) reciprocates to other leechers that send it pieces by giving higher priority to their requests. From the media server, BASS downloads pieces in-order, skipping over pieces that have already been downloaded by BitTorrent, or are currently in the process of being downloaded and are expected to finish before their playback deadline arrives. The system overview of BASS is given in figure 29. Even though BASS reduces the load at the server by a significant amount, the design of the system is still server oriented, and, hence, the bandwidth requirements at the server increase linearly with the number of users [98].

*Kangaroo* [101] is a system focused on providing both P2P VoD services as well as live streaming content. Kangaroo resembles a typical mesh-based P2P system in that it consists of *peers* coordinated by a *tracker*. Kangaroo handles DVD operations with minimum delays, network overhead, and server resources. Kangaroo implements a hybrid scheduling policy that combines selfish (sequential segment downloads for continuous playback) with altruistic (local rarest to improve segment diversity) behavior. A peer consists of several sub-components, the Segment Scheduler, the Peer Selection Scheduler, and the Neighborhood Manager. The Segment scheduler decides what segment should be scheduled for download next, while the Peer Selection Scheduler decides which neighbor peer(s) to schedule the download from and the Neighborhood manager that constantly re-visits the peer's neighborhood and decides which are the best peers to get/push data from/to. Each peer in Kangaroo downloads data in parallel from a small number of neighbors through data connections. Peers also maintain a number of control connections which are used to exchange information about available segments in a neighborhood, thus enabling the peer to infer the popularity and location of the segment for scheduling. Kangaroo resembles a gossip-based overlay in which a smart tracker is used to implement peer coordination. Kangaroo also provides low buffering times and high swarming throughput under user VCR-like operations. However, it doesn't consider user viewing behavior. Thus, VCR-like operations may cause long response time [102].

The *BiToS system* [103] is also based on BitTorrent. The main idea is to divide the missing blocks into two sets, "high priority set" and "remaining piece set", and request with higher probability blocks from the high priority set (Figure 30). The high priority set, contains all the pieces that are quite close to be reproduced. Thus, peer desires to download these pieces earlier, in contrast with the remaining pieces set, which contains pieces that won't be needed in the near future. After the initiation of the player, the Player Buffer requests the needed pieces from the received pieces buffer. In BiToS, the major emphasis is given for the careful scheduling of the video blocks. The pieces that miss their playback deadline are simply dropped. Hence, this may lead to degradation in video playback quality. Also due to asymmetric nature of the internet connections and heterogeneity of the peers, the system cannot guarantee that pieces requested are always available for playback on time [99].

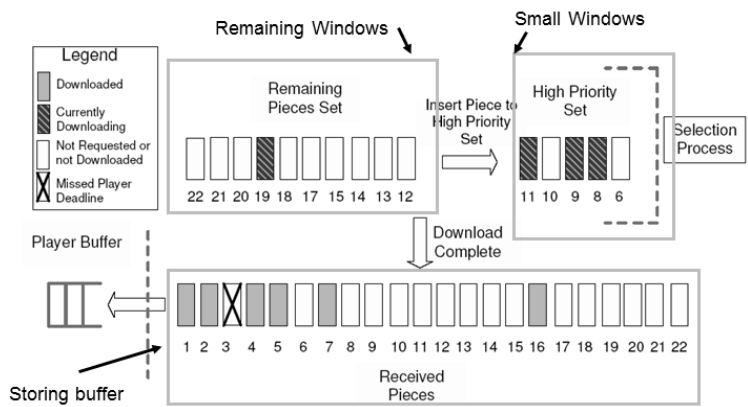

**Figure 30: BiToS Approach [103]**

*COCONET* [104] is a novel and efficient way of organizing peers to form an overlay network for supporting streaming and neighbor lookup for continuous playback or VCR operations. The scheme utilizes a cooperative cache based technique where each peer contributes a certain amount of storage to the system in return for receiving video blocks. The system uses this co-operative cache to organize the overlay network and serve peer requests, thereby reducing the server bottleneck supporting VCR related operations. In several P2P VoD systems, peers share video segments only with neighboring peers based on its playing position. However, in highly skewed viewing patterns, most of the peers are clustered around a particular playing position and very few peers are distributed at different positions throughout the video length. Hence, the peers may not find any or very few neighbors to satisfy their demand. COCONET avoids this situation. In order to find new supplier peers at different parts of the movie length, P2P VoD systems maintains an updated index of the live peers with their available video segments. Unlike other P2P VoD systems, COCONET does not use indexing at the tracker. Instead, the tracker only maintains a small subset of live peers which is queried only once as a rendezvous point when a new peer joins the system. Each COCONET peer builds an index based on the co-operative cache contents which helps to find any supplier peer for any video segment throughout the entire video length. The control overhead of COCONET is also low to maintain the overlay structure even during a heavy churn. COCONET also has better load balancing and fault tolerance properties. The distributed contributory storage caching scheme helps to spread the query load uniformly through the overlay and organizes the overlay in a uniform and randomized fashion which makes the content distribution independent from playing position.

## Mobile Video-on-Demand

Streaming video to mobile users is rapidly emerging as a crucial multimedia service. With the emergence of wireless technologies such as IEEE 802.11 and Bluetooth, mobile users are enabled to connect to each other directly without any networking infrastructure such as the Internet and infrastructure based wireless LANs. In other words, the users form a *mobile ad hoc network* (MANET). Due to the increasing popularity of wireless networks, mobile VoD systems have found many practical applications. For example, airlines can now provide VoD services in airport lounges to entertain the waiting passengers on their laptops or PDAs. Universities can install mobile VoD systems that allow students to watch important video lectures anywhere anytime on campus. In mobile VoD systems the equipments to watch the video broadcasting fall in a wide spectrum of heterogeneous capabilities, ranging from powerful laptops to primitive PDAs. One of main issue in mobile streaming is the heterogeneity found in mobile devices: diverse display size, computing power, memory, and media capabilities. The wireless bandwidth is limited whereas a video is typically large. A video server enabled with 802.11g could not deliver more than thirty six 1.5Mbps MPEG1 video streams at once to its wireless clients. However, 802.11b can only support at most seven concurrent such video streams. The load of a VoD system is usually distributed unevenly; it is heavy only over a short period of time. For instance, in the airport lounge example, the system would have a heavy load only during one or two hours before a flight departure. Therefore, the system should be able to adapt to different loads and make necessary adjustment to the broadcasting schedule so as to minimize the total bandwidth usage. Load adaptivity is also important to mobile VoD system because of its energy consumption. Since the coverage of wireless transmission is limited, we often need multiple hosts to cover a large enough service area. Due to this a significant amount of energy for the intermediate mobile hosts is consumed. In order to save energy, the system should use smaller total bandwidth when the load is light [105].

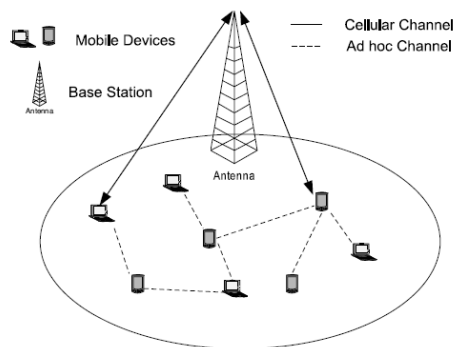

Figure 31: The unified cellular and ad-hoc network architecture

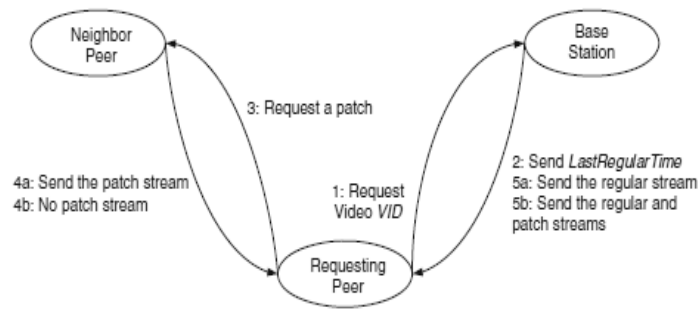

**Figure 32: Collaboration diagram of PatchPeer [106]**

*PatchPeer* [106] is a VoD technique for the wireless environment. The basic idea of PatchPeer is to take advantage of the distinct features of the hybrid wireless network (Figure 31) to overcome the scalability issue associated with the original Patching technique in a traditional wireless network. Figure 32 shows the typical interactions between a requesting peer with its neighboring peers and the server in PatchPeer. The requesting peer first sends the ID of the requested video to the server. It receives the starting time of the latest regular channel that is streaming the video from the server. The requesting peer then requests a patching stream from a neighboring peer to compensate for the initial missing part of the video. If a neighboring peer can provide the patching stream, the requesting peer receives the patching stream from the neighboring peer and the regular stream from the base station. Otherwise, the requesting peer receives both the patching stream and regular stream from the base station. The PatchPeer scales better than the original Patching, as most of the patching streams in PatchPeer are provided by mobile clients themselves, leaving the base station with more downlink bandwidth to serve more clients.

*MobiVoD* [107] is a mobile VoD system which employs a periodic broadcast protocol to achieve maximum scalability. The clients leverage an ad hoc network caching technique to minimize the service delay. The system consists of three components: *video server*, *clients*, and *local forwarders*. Due to the limitation of wireless transmissions, a video server cannot transmit a video to clients located in a wide geographic range. So, a scatter of stationary and dedicated computers called *local forwarders* is provided to relay the service to client's transmission coverage area. This area is called a local service area. If a client is within the service area of a local forwarder, the former can receive the video packets broadcast from the latter. The server and set of local forwarders form a service backbone. The service backbone is interconnected either via a wired WAN/LAN or via an infrastructure-based wireless network. A video is divided into segments, each broadcasts on a separate communication channel. When a new client joins the system, it waits until the next broadcast of the first segment starts to download the first segment. After playing the first segment, the client immediately switches to the broadcast of the second segment to download it, and so on until all segments have been downloaded. Periodic broadcasting makes the system scalable with increase in number of clients. However, as the period a new client must wait before it starts the VoD service is significant, MobiVoD employs two caching policies: Random-cache and Dominating-Set Cache (DSC). Random-cache permits a client to cache the first segment with some probability. Even if a new client finds some clients in its neighborhood, the chance of keeping cache by these clients may be low in random cache. Hence, as an alternative DSC which maintains a dominating set of the clients '$D_{set}$' is used. A client belonging to $D_{set}$ caches the first video segment. Using any one of the caching schemes, when a new client requests the VoD service, it joins the current broadcast immediately and downloads the video packets broadcast into a playback buffer. As for the beginning portion that was already transmitted by the current broadcast, the new client downloads and plays it immediately from a nearby cache. The new client switches to the playback buffer to play the rest of the video. MobiVoD focuses on popular videos, while PatchPeer handles videos with diverse popularity. Moreover, MobiVoD is using only the wireless local area networks, while PatchPeer operates in a hybrid wireless environment.

*MOVi* (Mobile Opportunistic Video-on-demand) [108] is a mobile P2P video-on-demand application based on ubiquitous WiFi enabled devices such as smartphones and Ultra Mobile PCs. MOVi addresses challenges such as limited wireless communication range, user mobility and variable user population density by exploiting the opportunistic mix use of downlink and direct P2P communication for improving the overall system throughput. It exploits sparingly distributed access points, user mobility, unstable channel conditions and population density to provide a high bitrate on-demand video streaming service. MOVi is comprised of two logical components: a mobile client node called MOVi Client (MC) and a network of servers known collectively as MOVi Server (MS) (Figure 33). MOVi Server maintains three key functions. It derives a connectivity map of link quality between MCs, schedules the direct transfer of content segments between pairs of MCs based on the map, and tracks the content delivery and caching status of all MCs within its domain. A MOVi Client maintains two key functions: serves as a temporal cache to help content diffusion inside the MOVi network, and acts as a channel state monitor by periodically observing link quality to its neighboring MCs and updates any changes to MS. The content is stored at a central repository and is fragmented into multiple equal segments prior to distribution. Each segment is normally mapped to several packets. Upon receiving requests from MCs, MS delivers content segments over the downlink path as well as schedules direct segment exchanges between MCs. If direct P2P communication is not possible with other MCs, the requested segment is delivered from MS to the MC via the access point path. An MC has no knowledge of which content segments reside in its neighboring MC and it simply waits for direct communication triggers from MS. Once the MC receives all segments that make up a video frame segment, the frame is handed to the media player. The player then decodes the frame for playout. If there are missing segments within yet to be playout video frame segment, MC sends immediate on-demand request to the MS to recover the missing segments. Neighboring peer discovery in MOVi is carried out by evaluating Signal Interference to Noise Ratio (SINR) value between MCs and the active duration of neighboring MCs. MOVi is able to increase the number of supported concurrent users two fold compared with unicast based on-demand video streaming, as well as reducing video start-up delay by half.

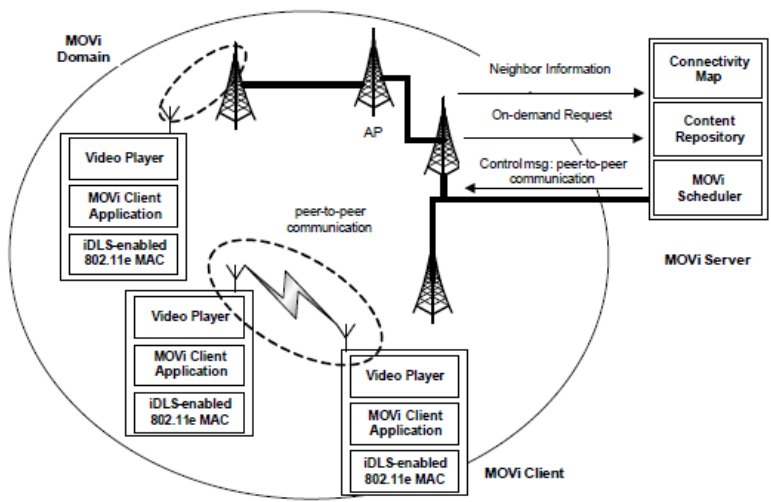

**Figure 33: MOVi Architectural Components [108]**

*Peer-to-Peer mobile video-on-demand (**P2MVOD**)* [109] allows a moving client to receive streaming data on demand from other moving clients in P2P architecture by utilizing multicast VoD technology. P2MVOD divides video content into the same sized segments. The segments are then broadcasted to eliminate the mobile routing overhead of unicast and multicast routing protocols. Segmenting the content enables multiple clients to share the accountability for providing all video content. A control server is used to control the segmentation of video content and the delivery of each segment. The control server does not store or deliver any video content. It only possesses information on the segments that can be provided by the clients that are storing them. Each client knows the control server address and submits a request for video delivery to it. On receiving a request, the server searches for a client with segments of the required content and forwards the request to that client. The control server holds and maintains a schedule that describes the time at which individual segments must be sent. By referring to the schedule, the control server determines the segments that must be sent to the new client and searches for other clients that can provide these segments. Further, it queries them regarding the possibility of sending at a time determined by it. The receiving clients do not need to know the identities of the clients. P2MVOD reduces the traffic on both the links compared to the patching technique, although it adds to traffic when the request rate is low.

# Mobile Live Streaming

Delivering media to large numbers of mobile users presents challenges due to the stringent requirements of streaming media, mobility, wireless, and scaling to support large numbers of users. Live streaming to mobile devices is thus a challenging task [110]. P2P based near-live video streaming is becoming more and more popular with users of fixed-line broadband network access, but it is mostly unavailable to mobile users.

In [111] presents a real-time P2P streaming system for the mobile environment. Peers are grouped into clusters according to their proximity in order to proficiently exchange data between peers. Clusters also help with scalability issues of peer maintenance. The architecture of the overlay network with three clusters sharing a certain streaming service is shown in Figure 34. Peers exchange actual media data between each other using RTP. RTP sessions are split into a number of partial streams in such a way that it allows reassembling the original media session in real-time at the receiving end. There is one Cluster Leader (CL) assigned to each cluster with the possibility for one or more Backup Cluster Leaders (BCLs). CLs are used to manage peers inside the cluster and to connect new arriving peers. Each ordinary peer performs periodical keep alive messaging to inform its existence to the CL and all other peers from which it has received RTP packets. This helps to avoid unnecessary data transmission because RTP uses User Datagram Protocol and the sending peer does not otherwise know that the receiving peer is still in the network. A new arriving peer selects a suitable cluster according to its best knowledge of locality using Round Trip Time values between CLs and itself. At the time of joining a cluster, a peer receives an initial list of peers from which the actual media data can be acquired. The corresponding CL inserts joined peers into its peer list. A peer finally selects its sources for the stream. When the cluster grows too large to be handled by a single CL, the cluster should be split into two separate clusters. The existing CL assigns one of its BCLs to become a new CL for the new cluster, and redirects a number of existing peers to the new cluster. Merging of two clusters must be done when a cluster becomes too small. All peers in the streaming network are forming a nonhierarchical mesh structure. The system offers very low initial buffering times.

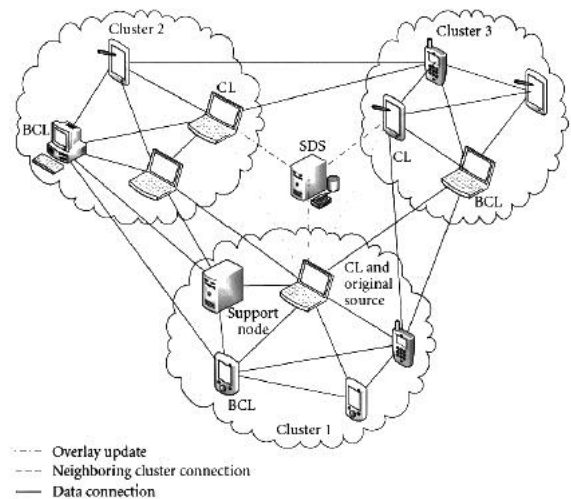

**Figure 34: Overlay architecture in a mobile environment [111]**

*LocalTree* [112] is a scalable algorithm which minimizes the energy used in packet transmission. Peers are organized into two tiers, the base tier and the tree tier. Peers are first connected in a simple unstructured mesh in the base tier. The base tier provides a network for further optimization of the energy consumption. In the base tier, peers utilize only local neighbor information to make independent distributed decisions on whether to rebroadcast a packet or not. The base-tier mesh is further optimized by the tree-tier algorithm. In the tree-tier, groups of relatively stable nodes are then identified based on node and link conditions. They are then connected following a greedy tree construction algorithm. With the two-tier operation, LocalTree is able to adapt different network dynamics.

# P2P Streaming and Cloud

Cloud Computing refers to both the applications delivered as services over the Internet and the hardware and systems software in the datacenters that provide those services. Cloud computing offers different service models as a base for successful end user applications. Due to the elastic infrastructure provided by the Cloud, it is suitable for delivering VoD and live video streaming services. A few schemes for media streaming which integrate the benefits of P2P and cloud technologies are proposed recently.

A streaming mechanism that merges P2P and cloud computing technologies for achieving efficient media streaming is proposed in [128]. Figure 35 represents an overall view of the proposed cloud-P2P architecture. The cloud contains multimedia streaming servers. The service has first level or directly connected clients ($C_1$, $C_2$, $C_3$, and $C_4$) and higher level clients ($HP_{11}$, $HP_{12}$ …). The first level clients after login consult and choose one among the three types of price packets. The price packets for customers are defined by considering three types of QoS parameters such as jitter, latency and bandwidth. Similarly, higher level clients acquire an information list with QoS status for all connected clients. The higher level clients contact the first or higher level customers instead of the provider for streams. There exists option for peers ($P_{111}$, $P_{112}$, $P_{121}$ ...) to connect to higher level clients who want to offer their service for free. The streaming network is thus organized into a P2P tree overlay. The service provider has direct centralized management for managing the contract policies among all types of customers.

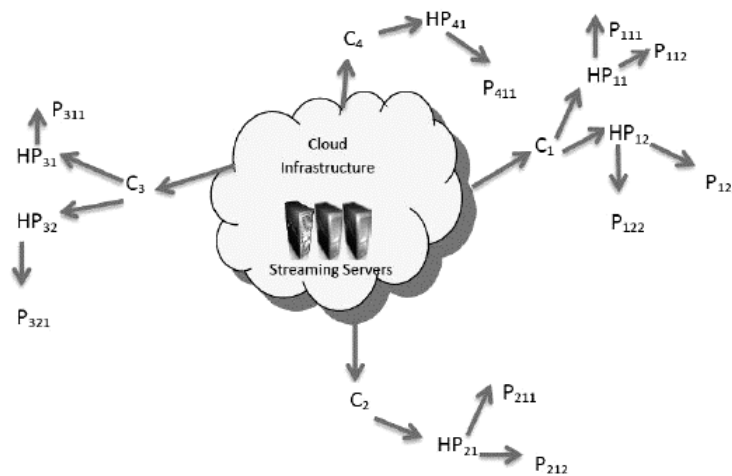

**Figure 35: P2P Streaming Cloud Architecture**

*AngelCast* [113] is a cloud-based live stream-acceleration service with optimized multi-tree construction that combines both P2P and cloud technologies. In P2P based live streaming systems the play out rates are constrained by the upload bandwidth of clients. Usually, the upload bandwidth is lower than download bandwidth for the participating peers. This limits the quality of the delivered stream. Therefore, to leverage P2P architectures without sacrificing the quality of the delivered stream, content providers use additional resources to complement those available through clients. In AngerlCast, a content provider is guaranteed that its clients would be able to download the stream at the desired rate without interruptions, while extremely utilizing the benefits from P2P delivery. AngelCast achieves this by employing special servers from the cloud, called *angels*. The angels can supplement the gap between the average client upload capacity and the desirable stream bit-rate. Angels download only the minimum fraction of the stream that enables them to fully utilize their upload bandwidth.

Figure 36 shows the architectural elements of AngelCast. The Registrar collects information about clients, making fast membership management decisions that ensure smooth streaming. When a new node joins the stream, it contacts the Registrar and informs it of its available upload bandwidth. The Registrar uses a data structure representing the streaming trees and assigns the new client to a parent node in each tree. The Registrar also decides how many future children the new node can adopt in each tree. Content providers contact the Registrar to enroll their streams. The Registrar uses the profiler to estimates the uplink capacity of clients. The Accountant uses the estimated gap between the clients' uplink capacity and the stream play out bit-rate to give the content provider an estimate of how many angels it will need. The service of angels' is utilized to achieve quality of service. AngelCast makes sure that any parent has at least two children except for the nodes in the second to last level, insuring a logarithmic depth of all trees.

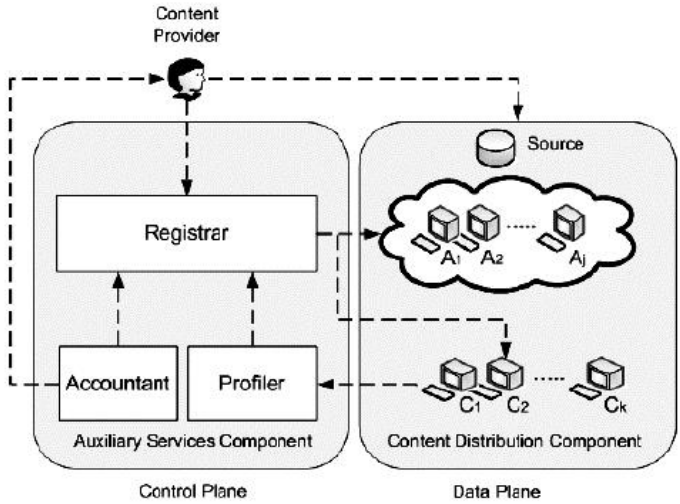

**Figure 36: The architectural elements of AngelCast** [113]

# Security in P2P Video Streaming

The open and anonymous nature of P2P network makes it an ideal medium for attackers to spread malicious content. As a result, widespread and unrestricted deployment of P2P systems exposed a number of security vulnerabilities. In a P2P environment the collaboration of all peers is very important for the correct functioning of the system. Every peer is soaking up network bandwidth. If too many users access the same network resource, the network bandwidth may be used up, resulting in a denial of service. A malicious node would continuously issue queries with high Time-To-Live (TTL) values on the network, thus generating huge amount of network traffic rendering the network unusable by other honest peers. The peer who offers a resource may go offline while other fellow peers are downloading from it. A malicious peer may just simply route a query to a non-existent peer or an unreliable peer with long latency.

Security-wise, P2P streaming systems are more challenging than other P2P applications because they are more vulnerable to QoS fluctuations. Live streaming protocols are most sensitive to delay and delay jitter. If a user is not receiving packets in time, he/she may grow dissatisfied with the quality of the delivery and leave the system altogether. Due to this, the peers connected to that machine may also be affected. Even minor quality variations cause the viewing experience to loose appeal and the user to drop the service. P2P streaming is vulnerable to manipulation and threats at the transport and network layers. Clever attacks can compromise selectively the guarantees that a streaming session should provide, rendering some channels unusable, or making the broadcast unavailable in particular locations [114].

Traditional security mechanisms typically protect resources from malicious users, by restricting access to only authorized users. However, the problems in P2P systems relate more with trustworthiness rather than security. Therefore, there is demand for mechanisms to maintain the trust of P2P systems. Trust management is a successful approach that helps to maintain overall credibility level of the system as well as to encourage honest and cooperative behavior. The inspiration of trust management is that since in a P2P system there is no central authority that can authenticate and guard against the actions of malicious peers, it is up to the peer to protect itself and to be responsible for its own actions. Consequently, each peer in the system needs to somehow assess information received from another peer in order to determine the trustworthiness of both the information as well as the sender. This can be attained in many ways such as relying on direct experiences or obtaining reputation information from other peers [115].

*Common Attacks and Solutions*

*Denial of Service (DoS) attacks:* Denial of Service attacks decrease or cease total capable network activity. The goal of such an attack is to exhaust key resources at the target, diminishing the target's capacity to either provide or receive service. Resources that can be exhausted include the target's downstream bandwidth, upstream bandwidth, CPU processing, or TCP connection resources. Compared with the widely used file-sharing networks, P2P streaming networks are more vulnerable to DoS attacks. There are several reasons for DoS attacks in P2P streaming. Video streaming requires high bandwidth. Hence, a certain amount of data loss could make the whole stream useless. Since streaming applications require their data to be delivered in a timely fashion, data with a missed deadline are useless. Usually a streaming network consists of a limited number of data sources. Hence, the failure of the data source could bring down the whole streaming system [116]. For example, malicious nodes send excessive amounts of requests or duplicate packets intended for their peers. Thus, a fair node would be flooded with useless messages or too many requests for it to handle. Consequently, the ability to bring a contribution to the streaming session is compromised. In this way, the resources of the system are exhausted with a relatively small effort on the attacker side.

*Ripple-stream* [116] is a DoS resilience framework which employs a credit system to allow peers to evaluate other peers' behaviors. The overlay is organized according to a credit-constrained peer selection mechanism. The peers share the credit information with each other. Peers with high trustworthiness are kept in the central part of the overlay structure. Malicious nodes, with low reliability, are pushed to the peripheral of the network. The higher the credit, the closer a peer can be to the data source. Credit management component in the system translates a user's behavior to its credit value. In Ripple-stream, when a new peer A joins the overlay, it will first obtain a list of peers with mediocre credit from a bootstrap mechanism. After joining the overlay, A accumulates credit by fulfilling its duties. This credit related operations are handled by the credit component included in ripple-stream. In the meantime, A also tries to find upstream peers that can provide better service based on some overlay optimization principles. If A discovers malicious behaviors of other peers, it disconnects from these peers and reports its discovery to the credit system. An example of a ripple-stream based overlay is shown in figure 37. Ripple-stream achieves DoS resilience with the credit system and during attacks the ripple-stream stabilizes the overlay and substantially improves the streaming quality.

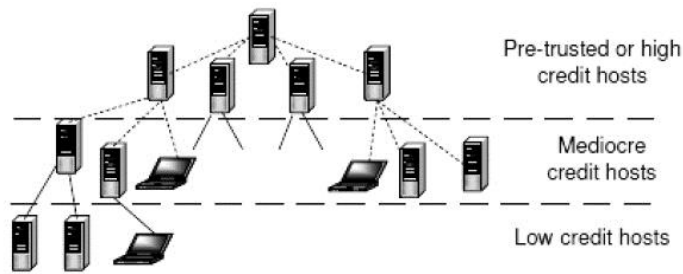

**Figure 37: Example of a ripple-stream based overlay**

*Free-riding:* Nodes that consume services offered by other nodes but which do not themselves contribute services to the P2P network are known as free-riders. In P2P networks, free-riding is a familiar problem. A free-rider guzzles more resources than it contributes. In the case of P2P streaming, a free-rider is a peer who downloads data but uploads little or no data in return. The encumbrance of uploading is on the unselfish peers, who may be too few in number to provide all peers with a satisfactory quality of service. In both live streaming and VoD, peers require a minimal download speed to sustain playback. Hence, free-riding is very harmful as the unselfish peers alone may not be able to provide all the peers with sufficient download speeds.

A classification of free riding techniques for file sharing applications is presented in [117]. The schemes are categorized as monetary-, reciprocity-, and reputation-based approaches. Monetary-based approaches charge peers for the services they receive. Because these services are still very low cost, such approaches are also called micropayment-based solutions. The technique proposed in [118] is an example for monetary-based approach. The main disadvantage is that the proposed solutions require some centralized authority to monitor each peer's balance and transactions. This can cause scalability and single-point-of-failure problems. In reciprocity-based approaches, a peer monitors other peers' behaviors and evaluates their contribution levels. The well-known P2P application BitTorrent implements a reciprocity based approach by adjusting a peer's download speed according to its upload speed. Reciprocity-based approaches face several implementation issues such as fake services published by peers. Since peer itself provides contribution level information, the credibility is in question. In reputation-based approaches peers with good reputations are offered better services. These approaches construct reputation information about a peer on the basis of feedback from other peers. Reputation-based approaches store and manage long-term peer histories. XRep [119] is an example of an autonomous reputation system. Reputation sharing is achieved in XRep through a distributed algorithm by which resource requestors can evaluate the consistency of a resource offered by a participant before beginning the download.

In [120] proposes two policies to limit the number of free-riders in a P2P streaming system - *Block and Drop* and the *Block and Wait policies*. With the Block and Drop (BD) policy, free-riders that would like to join the streaming session are blocked if the free upload capacity in the overlay is less than the streaming rate. Under the Block and Wait (BW) policy, free-riders are blocked if the overlay does not have enough available upload capacity. The same users can be temporarily disconnected and have to wait to reconnect if there is not enough capacity to serve all peers. Under the BD policy both the blocking and the dropping probabilities can be high. Therefore, the free-riders already admitted to the system are frequently dropped. Under the BW policy, the number of free-riders waiting to be reconnected is very low for all parameter settings. As a result, free-riders receive the stream without interruption with high probability. This feature makes the BW policy a good option to control free-riders.

In [121] proposes a mechanism for Give-to-Get free-riding-resilient for P2P VoD systems. In Give-to-Get, peers have to forward the chunks received from a peer to others in order to get more chunks from that peer. By preferring to serve good forwarders, free-riders are excluded in favour of well-behaving peers. When bandwidth in the P2P system becomes scarce, the free-riders will experience a significant drop in the experienced quality of service. Free-riders will thus be able to obtain video data only if there is spare capacity in the system.

In [122] proposes a rank-based peer-selection mechanism for peer-to-peer media streaming systems. The mechanism provides incentives for cooperation through service differentiation. Contributors to the system are rewarded with flexibility and choice in peer selection to provide high quality streaming sessions. Free-riders are given limited options in peer selection and hence receive low quality streaming. The contribution of a user is converted into a score, then the score is mapped into a rank, and the rank provides flexibility in peer selection. Cooperative users earn higher rank by contributing their resources to others, and eventually receive high quality streaming. Free riders have limited choice in peer selection, hence receive low quality streaming. The incentive mechanism reduces the data redundancy required during a streaming session to tolerate packet loss.

A payment-based incentive mechanism for P2P live media streaming is proposed in [123]. In a payment-based system, the P2P network is treated as a market. Every overlay node plays the double role of service consumer and provider. Consumers try to buy the best possible service from service providers at a minimum price, while the providers strategically decide their respective prices in a pricing game, in order to maximize their economic revenues in the long run. A peer earns points by forwarding data to others. The data streaming is divided into fixed length periods, during each of which peers compete with each other for good data suppliers for the next period in a first-price auction like procedure using their points. Once a peer finds an ideal parent, it takes part in a competition for that parent. If it wins, it becomes a child of that parent; otherwise it gets a list of the winner peers, from which it attempts to find a new best parent. It again takes part in the competition for that new parent and continues this process until it wins a parent or has no parents to choose. In the latter case it tries to find a parent in a best effort manner.

*Pollution attacks:* In a P2P live video streaming system, a polluter can introduce corrupted chunks. Explicitly, an attacker can join a current video channel and create partnerships with other peers watching the channel. The attacker can then announce to its partners that it has a large number of chunks for the current video stream. When the neighbors request advertised chunks, the attacker sends bogus polluted chunks in place of legitimate chunks. Each receiver mixes into its playback stream the polluted chunks it receives from the attacker along with other chunks it receives from its other neighbors. The polluted chunks damage the quality of the rendered video at the receiver. Polluted chunks received by an unsuspicious peer not only effect that single peer, but since the peer also forwards chunks to other peers, and those peers in turn forward chunks to more peers, and so on, the polluted content can potentially spread through much of the P2P network. If the amount of polluted data is very important, users might ultimately get unsatisfied and completely stop using the system [125]. Figure 38 shows [124] a P2P network where a polluter sends bogus chunks to other peers, falsely marking these chunks as legitimate. These corrupted chunks get propagated through the P2P network. Content pollution attacks can severely impact the quality-of-service in P2P live video streaming systems. There are solutions proposed in the literature for managing the pollution attacks in P2P file sharing applications. However, few schemes are available for fighting pollution attacks in P2P streaming applications.

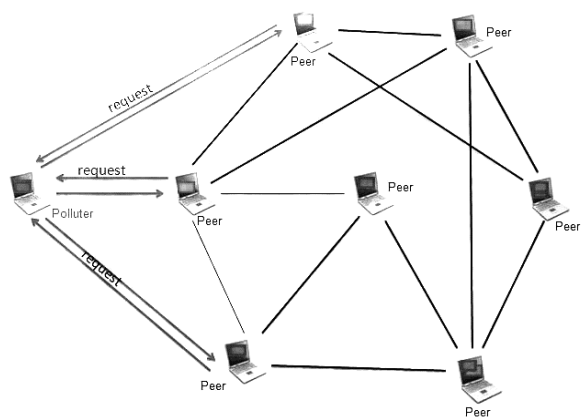

**Figure 38: Pollution Attack in a P2P Live Video Streaming System**

In [126] introduced a light-weight non-repudiation protocol called Malicious node Identification Scheme (MIS) for network-coding based P2P Streaming Networks. With a network coding technique, instead of merely relaying the packets of information they receive, the nodes of a network will take several packets and combine them together for transmission. However, the "combination" nature of network coding makes it vulnerable to pollution attacks. MIS employs an approach for detecting the existence of malicious nodes. Each decoding node detect corrupted blocks by checking if the decoding result matches the specific formats of video streams. Any node having an unreliable decoding result will send an alert to the servers to trigger the process of recognizing malicious nodes. The servers then compute a checksum based on the original blocks, and distribute it to the nodes using the streaming overlay. The checksum helps the nodes to detect which neighbor has sent it a corrupted block. The accuracy of MIS is based on the condition that no node can lie when reporting a suspicious node that has sent a corrupted block, and evidence associated with the corrupted block is necessary to demonstrate to the servers that the reported node has really sent the block. A non-repudiation transmission protocol is used to achieve this. MIS has good computational efficiency and the ability of managing a large number of corrupted blocks and malicious nodes. However, this scheme requires a distribution of multiple checksums to all the peers when an attack is detected, which incurs significant communication overhead. In [127] proposes a trust management system that identifies attackers and excludes them from further sharing of multimedia data to resist pollution attacks in P2P live streaming.

## Summary

Video-over-IP applications have recently attracted a large number of users on the Internet. With streaming, a user does not have to wait to download a file to play it and can watch the video in real time. The basic solution for streaming video over the Internet is the traditional client-server service model. A client sets up a connection with a video source and video content is streamed to the client directly from the server. However, the client-server design, harshly restrict the number of concurrent users in video streaming due to the bandwidth bottleneck at the server side. Another model, content delivery networks overwhelmed the same bottleneck issue by adding dedicated servers at physically dissimilar locations. This results in expensive deployment and maintenance.

P2P networking is a very promising model to construct various distributed applications. Recently, quite a few P2P streaming systems have been deployed to provide live and VoD streaming services on the Internet. Compared to conventional approaches, the main benefit of P2P streaming is that each peer contributes its own resources to the streaming session. Administration, maintenance, and responsibility for operations are hence dispersed among several users instead of focusing on few servers. Due to this, there is a rise in the quantity of resources in the network. Accordingly, the usual bottleneck problem of the client-server systems is further reduced. The P2P architecture thus extends extremely well with large user population, and also provides a scalable and economical alternate to traditional streaming services. Mainly there are two well-known schemes for P2P video streaming: tree- and mesh-based. A hybrid of the two schemes is also emerging. The chapter reviewed the architecture of these models and also briefly explained few systems built-on these approaches. The application of P2P technology in video streaming on MANET is known as Mobile P2P streaming. The issue of live video streaming on MANET is still a real challenge due to frequent changes in network topology, and the sensitiveness of radio links. A few schemes for media streaming which integrate the benefits of P2P and cloud technologies are proposed recently. Security has significant impact on P2P based streaming applications. Media streaming is inherently more prone to attacks as it is very difficult to monitor the participating peers in the overlay. The network consists of thousands of nodes, not all can be trusted. Security forms one of the most critical issues in a streaming system. The chapter also reviewed various security issues and mechanisms for preventing such attacks.